\documentclass[10pt,final]{IEEEtran}
\usepackage{amsmath}
\usepackage{amssymb}
\usepackage{amsfonts}
\usepackage{latexsym}
\usepackage{graphicx}
\usepackage{enumerate}
\usepackage{multicol}
\usepackage{color}
\usepackage{array}
\usepackage{algorithm,algorithmic}
\usepackage{bm}
\IEEEoverridecommandlockouts
\allowdisplaybreaks[0]

\begin{document}
\title{Design of RIS-UAV-Assisted LEO Satellite Constellation Communication}

\author{Wenfei Yao, Xiaoming Chen, Qi Wang, and Xingyu Peng

\thanks{Wenfei Yao, Xiaoming Chen, Qi Wang, and Xingyu Peng are with the College of Information Science and Electronic Engineering, Zhejiang University, Hangzhou 310027, China (e-mails: \{yaowenfei, chen\underline{~}xiaoming, wang-qi, peng\underline{~}xingyu\}@zju.edu.cn).}}
\maketitle

\begin{abstract}
Low Earth orbit (LEO) satellite constellations play a pivotal role in sixth-generation (6G) wireless networks by providing global coverage, massive connections, and huge capacity. In this paper, we present a novel LEO satellite constellation communication framework, where a reconfigurable intelligent surface-mounted unmanned aerial vehicle (RIS-UAV) is deployed to improve the communication quality of multiple terrestrial user equipments (UEs) under the condition of long distance between satellite and ground. To reduce the overhead for channel state information (CSI) acquisition with multiple-satellite collaboration, statistical CSI (sCSI) is utilized in the system. In such a situation, we first derive an approximated but exact expression for ergodic rate of each UE. Then, we aim to maximize the minimum approximated UE ergodic rate by the proposed alternating optimization (AO)-based algorithm that jointly optimizes LEO satellite beamforming, RIS phase shift, and UAV trajectory. Finally, extensive simulations are conducted to demonstrate the superiority of the proposed algorithm in terms of spectrum efficiency over baseline algorithms.
\end{abstract}


\section{Introduction}
The sixth generation (6G) wireless networks will enable global coverage and massive communication \cite{6G}. In this vision, low Earth orbit (LEO) satellite constellation plays a crucial role in 6G wireless networks. LEO satellite communications have shorter signal propagation delay due to low orbital altitude, and thus satisfy delay requirements of most wireless services. Meanwhile, the global coverage and inter-satellite communications capability of the LEO satellite constellation enable 6G networks to extend to remote areas that are hard to reach by traditional terrestrial networks, realizing true global interconnection \cite{LEO satellite constellation}. Given the potential promise of LEO satellite constellations, several projects, represented by Starlink, OneWeb, Kuiper, and Telesat, are being actively pursued and have made significant progress \cite{LEO satellite constellation project}.

In order to meet the requirements of higher connection density and increasing broadband data traffic with limited spectrum, multi-beam satellite systems (MSSs) with aggressive full-frequency reuse (FFR) modes have been investigated in satellite communications \cite{ChuJianHang}-\cite{linear precoding for massive MIMO}. For the inter-beam co-channel interference (IBCCI) caused by FFR, some interference management techniques have been proposed, such as frame-based precoding \cite{frame-based precoding} and linear precoding with massive antenna array \cite{linear precoding for massive MIMO}, which efficiently utilize the bandwidth with separated beams while ensuring low IBCCI. Further, multi-satellite collaboration technology with MSS has been investigated under the context of LEO satellite constellation \cite{LEO satellite cluster}-\cite{Dynamic Resource Allocation}. In particular, the work \cite{LEO satellite cluster} introduced the concept of satellite clusters, which enhance the performance of non-terrestrial networks through cooperative transmissions among closely located satellites. For the collaboration among the satellites, a leader-follower structure is then proposed, whereby the leader satellite manages the entire cluster and exchanges information and control signals with the follower satellites via inter-satellite links (ISLs). The authors in \cite{Coordinated precoding} investigated the coordinated precoding using statistical channel state information (CSI) for multi-satellite systems, where instantaneous CSI can not be entirely and timely exchanged between satellites. This approach significantly reduces the overhead of information interaction between LEO satellites. Moreover, the work in \cite{Multi-satellite cooperative networks} proposed a low-complexity framework for multi-satellite cooperation in LEO constellation networks that enhanced spectral efficiency through joint beamforming and user scheduling. These studies have shown that multi-satellite collaboration technology can significantly increase system performance and mitigate link budget shortfalls due to limited transmit power budget of individual satellites.

Nevertheless, in a line-of-sight (LoS)-dominated satellite-terrestrial propagation environment of FFR MSSs, it is difficult to eliminate IBCCI with precoding techniques when user equipments (UEs) are densely distributed \cite{hotspot}. As a result, FFR MSSs cannot fully meet the on-demand and real-time capacity requirements of high connection density UEs. To address this issue, a satellite-aerial-terrestrial integrated network (SATIN) has been investigated by employing unmanned aerial vehicles (UAVs) as relays \cite{SATIN}. UAVs can fly at much lower altitudes than LEO satellites and serve UEs effectively due to their maneuverability, flexibility, and ease of deployment. Besides, the LoS-dominated satellite-terrestrial channel is replaced by a shorter-range LoS-dominated channel, which can significantly improve the system performance \cite{3D Channel Tracking for UAV-Satellite Communications}. For instance, the authors in \cite{Joint Resource Allocation and UAV Trajectory Optimization} proposed an iterative algorithm that significantly enhanced the system capacity by jointly designing smart device connection scheduling, power control, and UAV trajectory. It is worth noting that, in a UAV-assisted system, the long enduration and high reliability of the UAV are the key factors in ensuring system performance, enabling the UAV to carry out more sustained missions and enhancing the coherence and coverage of mission execution. To this end, the work in \cite{UAV-LEO Integrated Backbone} investigated the problem of maximizing the uploaded data rate to LEO satellites and minimizing the energy consumption simultaneously. By optimizing the bandwidth allocation and UAVs trajectory, the data gathering efficiency is significantly improved, while the total energy consumption of UAVs is reduced by optimizing the transmit power of UAVs and the selection of LEO satellites. Considering limited payloads of UAVs, the work in \cite{LEO-Satellite-Assisted UAV} proposed a column generation-based algorithm to minimize the total energy cost of UAVs while satisfying the internet of remote things demands in the SATIN. Further, to satisfy the quality of service of UEs and maximize the energy efficiency, the authors in \cite{Joint Resources Allocation and 3D Trajectory Optimization} proposed an effective algorithm by jointly optimizing the UAV 3D trajectory and resource allocation. These studies suggest that deploying UAVs as relay in SATIN to improve performance is a viable approach, but ignore the additional computational resources required as well as the spectrum resources. Besides, the power consumption of UAV for forwarding signals is still a challenge for UAVs sustainability \cite{A Survey of Air-to-Ground Propagation Channel}.

Recently, reconfigurable intelligent surface (RIS)-mounted UAV (RIS-UAV) has attracted increasing interest. Specifically, RIS is a two-dimensional array plane consisting of a large number of passive elements, which can transmit or reflect the incident signal with a specific phase shift \cite{RIS definition}. In addition to deployment flexibility, it requires less complex hardware than traditional relay modes, such as amplify-and-forward (AF) or decode-and-forward (DF) \cite{On the design of RIS-UAV}. Moreover, the integration of RIS and UAVs also facilitates the use of high-frequency bands to create customizable wireless environments and enhance system performance. For instance, the work \cite{Phase shift design for RIS-assisted SATIN} introduced RIS into SATIN and optimized the phase shift of the RIS by maximizing the signal-to-interference-plus-noise ratio (SINR) of the UE in the presence of interfering satellite. However, due to the high maneuverability of UAVs, obtaining the accurate CSI about the channel from the RIS to the UEs in real-time is challenging. To this end, the authors in \cite{Secure communication in UAV-RIS-empowered multiuser networks} presented a joint optimization framework for beamforming, phase shift, and UAV trajectory in UAV-mounted RIS-empowered networks to enhance secure communication, taking into account channel uncertainty. Furthermore, the paper \cite{UAV-RIS-aided SATIN} presented a UAV-RIS assisted interference alignment design for SATIN with different types of CSI, i.e., no CSI, statistical CSI, and delayed CSI. By designing interference management, beamforming, and space-coding at the satellite and introducing RIS-UAV for the cooperating interference elimination process, the system capacity was significantly improved. In addition to imperfect CSI, hardware impairment is a key factor affecting the performance of the RIS-assisted system. Hardware impairments critically degrade RIS-aided system performance by introducing phase noise, signal distortion, and estimation errors \cite{HWI1}-\cite{HWI2}. To address the influence of hardware impairment, some works have been investigated. For instance, the authors in \cite{HWI3} focused on the transceiver hardware impairment and imperfect CSI, and formulated the linear minimum mean square error estimator of the equivalent RIS-assisted channel. The work \cite{HWI4} theoretically formulated the ergodic sum rate of the STAR-RIS assisted non-orthogonal multiple access (NOMA) uplink while taking channel estimation errors and hardware impairments into consideration. In \cite{HWI5}, the authors derived the optimal receive combination and transmit beamforming vectors, and provided analytical upper and lower bounds for maximum energy efficiency.

Although the aforementioned studies have shown that integrating RIS-UAV with SATIN may be effective in improving performance, most of them have only considered the simple case of a single satellite. In future LEO satellite constellations, inter-satellite collaboration is a key technique for communication performance improvement. Moreover, the channel refinement capability of RIS for LoS-dominated satellite-terrestrial propagation environments in LEO satellite constellations with MSS remains unrevealed. Therefore, how to introduce RIS-UAV to the LEO satellite constellation communication is still an open issue. Based on the above considerations, a RIS-UAV-assisted LEO satellite constellation communication framework is investigated in this paper. Specifically, a RIS-UAV is deployed in the system to enhance the satellite-ground communications by introducing RIS-assisted links in a LoS-dominated satellite-terrestrial environment. In order to achieve communication fairness between UEs, we formulate the optimization problem of maximizing the minimum UE ergodic rate by jointly designing LEO satellite beamforming, RIS phase shift, and RIS-UAV trajectory. The major contributions of our work are listed below:
\begin{enumerate}

    \item We present a novel RIS-UAV-assisted LEO satellite constellation communication framework, which leverages the collaboration between LEO satellites and introduces RIS-UAV to enhance multiuser communication quality in the LoS-dominated satellite-terrestrial environment.

    \item To reduce the pilot overhead, the approximated ergodic rate of each UE is derived. Building on this, an alternating optimization (AO)-based algorithm is proposed to ensure fairness by maximizing the minimum ergodic rate among UEs, through the joint design of LEO satellite beamforming, RIS phase shift, and RIS-UAV trajectory.

    \item Both theoretical analysis and numerical simulations validate the superiority of the proposed algorithm, demonstrating significant performance improvements over existing benchmark algorithms as well as existing related works.
\end{enumerate}

\subsection{Organization and Notations}
The subsequent sections of this paper are structured as follows. Section II introduces the system model of RIS-UAV-assisted LEO satellite constellation communication. Then, Section III formulates the optimization problem and proposes a joint design algorithm for RIS-UAV-assisted LEO satellite constellation communication. Section IV validates the effectiveness of the proposed AO-based algorithm through simulation results. Lastly, Section V concludes the paper.

\emph{Notations}: $\mathbb{C}^{M\times N}$ denotes a complex matrix of size $M \times N$.  Bold uppercase and lowercase letters denote matrices and column vectors, respectively. $(\cdot)^{\text{T}}$ and $(\cdot)^\text{H}$ denote the transpose and conjugate transpose, respectively. $\odot$ represents the Hadamard product. $\otimes$ denotes the Kronecker product. $\mathcal{CN}(\mu,\sigma^2)$ denotes the circularly symmetric complex Gaussian distribution with $\mu$ being the mean and $\sigma^2$ being the variance. $|\cdot|$ and $\|\cdot\|$ denote a scalar's absolute value and a vector's L2-norm, respectively. $\text{diag}(\cdot)$ and $\text{vec}(\cdot)$ denote the process of diagonalization and vectorization, respectively. $\mathbf{J}_{N \times M}$ and $\mathbf{I}_{N}$ denote the matrix of size $N \times M$ whose elements are all 1 and the unit matrix of size $N \times N$, respectively. $\mathbf{C}_{m,m}$ denotes the $m$-th diagonal element of the matrix $\mathbf{C}$.

\section{System model}

\begin{figure}[t]
	\centering
	\includegraphics[width=3.4in]{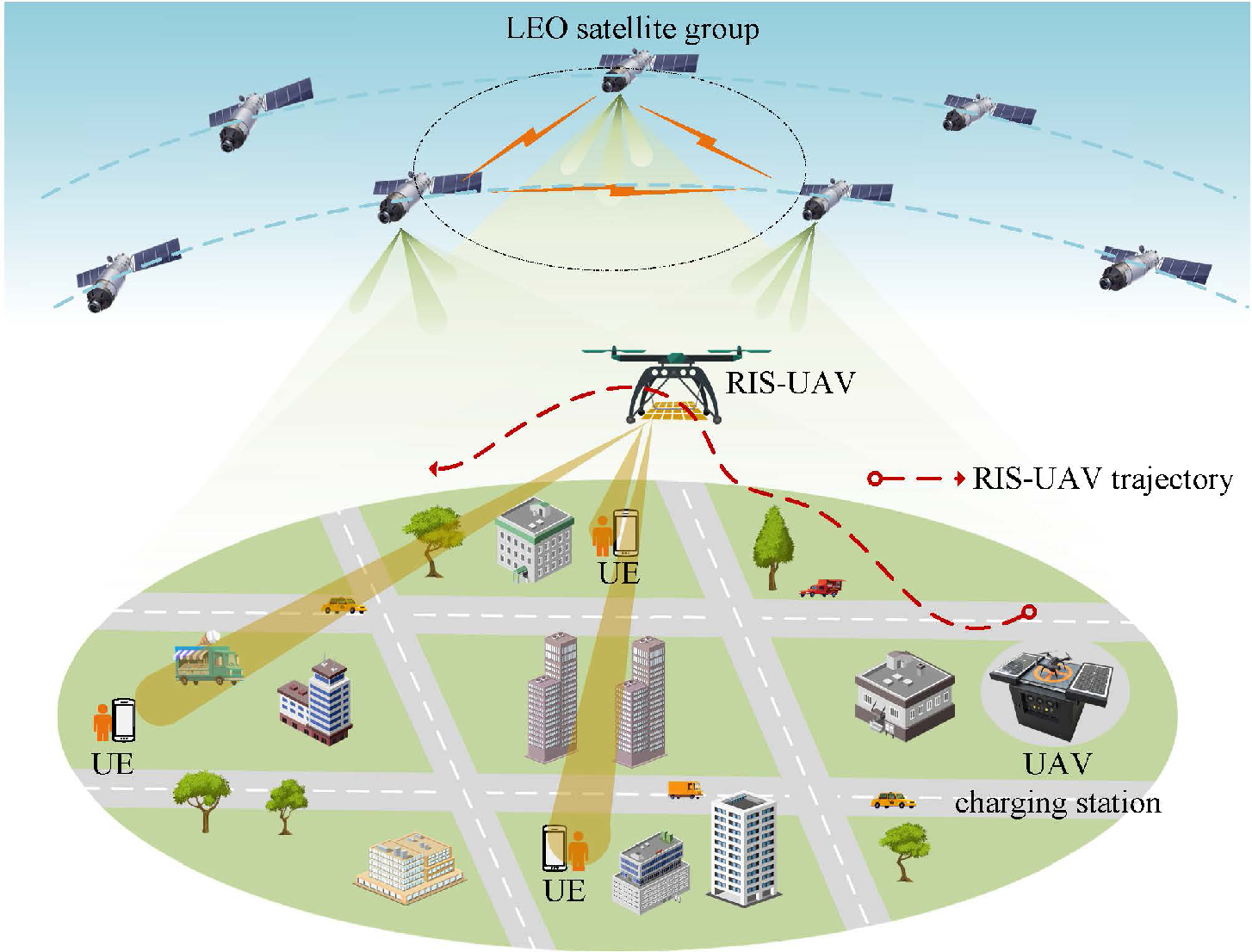}
	\caption{System model of the RIS-UAV-assisted LEO satellite constellation communication}
	\label{system model}
\end{figure}

Consider a RIS-UAV-assisted LEO satellite constellation communication system, as shown in Fig. \ref{system model}, where a large number of LEO satellites provide communication services to terrestrial UEs. In particular, the LEO satellite is configured with a satellite pose control system to keep its $N_s$-antenna uniform planner array (UPA) oriented towards the center of the Earth. In a specific region, a total of $K$ single-antenna UEs are covered and served by a group of $S$ visible LEO satellites over a time-frequency resource block. To ensure quality of service, the number of UEs $K$ is limited to no more than the number of satellite antennas by demand-based user scheduling\footnote{The user scheduling strategy employs demand-driven proportional fair scheduling, where UEs are prioritized based on their unmet demand and channel condition. Specifically, the unmet demand ratio of the $k$-th UEs is defined as $\alpha_{k} = \frac{Z_k}{D_k}$, where $Z_k$ is the remaining demand and $D_k$ is the total requested demand. Based on this, the first scheduled UE is the one maximizing the metric $\alpha_{k}\Vert \mathbf{f}_{k} \Vert$, where $\mathbf{f}_{k}$ denotes the channel from the LEO satellite group to the UE. For the remaining UEs, the scheduling metric for the $m$-th UE is defined as $w_{m} = \alpha_{m}(1-\sum\limits_{j \in \varkappa}\frac{\vert\mathbf{f}_{j}\mathbf{f}_{m}^{\text{H}}\vert}{\Vert\mathbf{f}_{j}\Vert\Vert\mathbf{f}_{m}\Vert})$, where $\varkappa$ denotes the set of already scheduled UEs. The UE with maximum $w_{m}$ is selected next, achieving a balance between priority and orthogonality to previously scheduled UEs \cite{user scheduling}.}. ISLs for information exchange exist among these LEO satellites. Considering large propagation loss of LEO satellite communications, a UAV flying at a fixed height of $h_0$ is employed in the region to enhance the downlink communication services, where a RIS with $M_r$-elements is deployed flat below the UAV. The system adopts a leader-follower structure for the collaborative satellites where a dynamically selected leader satellite collects full sCSI, executes the joint optimization, and finally transmits optimized parameters to the follower satellites and the RIS-UAV via dedicated control links\footnote{We assume that the overheads and transmission delays brought by wireless control are negligible due to the small size of optimized parameter data and the fast transmission rate of the dedicated control channel.} \cite{LEO satellite cluster}.

A data frame of duration $T$ contains $N$ time slots, where the slot duration $\delta=T/N$ is carefully chosen to balance performance and overhead. In particular, the slot duration $\delta$ is chosen to be sufficiently short so that RIS-UAV and UE positions can be considered static within each time slot, yet long enough to avoid excessive computational and control overhead due to frequent updates. The communication system is built on a Cartesian coordinated system in which the UAV charging station is set as the coordinate origin $\emph{O} = [0,0,0]$, and the $x$-axis and $y$-axis are set to be parallel to the lines of latitude and longitude, respectively. In general, the coordinates of RIS-UAVs and UEs can be accurately acquired through the global navigation satellite system (GNSS) and then fed back to the leader satellite \cite{3GPP}. Without loss of generality, the positions of the RIS-UAV and $k$-th UEs within the $n$-th time slot are defined as $\mathbf{q}^{r}[n] = [x[n],y[n],h_0]$ and $\mathbf{q}_{k}^{u}[n] = [x_k[n],y_k[n],z_k[n]]$, respectively. Accordingly, the trajectory of RIS-UAV and UEs follows the time-varying coordinates of $\{\mathbf{q}^{r}[n]\}_{n=1}^{N}$ and $\{\mathbf{q}^{u}_{k}[n]\}_{n=1}^{N}$, respectively. Note that the movement of RIS-UAV between two adjacent time slots is limited by the maximum speed $V_{\max}$. Moreover, the RIS-UAV has a limited flight range due to its limited energy and the need to prevent collision and interference with RIS-UAVs in other regions. To sum up, the movement of RIS-UAV should be subject to the following constraints
\begin{align}\label{RIS-UAV Trajectory Constraints}
    &\mathbf{q}^r[0] = [0,0,h_0], \\
    &\Vert \mathbf{q}^r[n] - \mathbf{q}^r[n-1] \Vert^2 \leq \delta V_{\max}, \\
    & \Vert\mathbf{q}^r[n] \Vert \leq l_{\max},
\end{align}
where $l_{\max}$ represents the maximum flight radius of the RIS-UAV. In addition, there is a constraint for the UAV's energy consumption. In this paper, we only consider the power consumption for the UAV propulsion since the RIS and the UAV communication consume very little energy \cite{RIS_power}. According to the \cite{UAV_power}, the propulsion power of the UAV can be formulated as
\begin{align}
    P(V) = P_{b} +  P_{i} +  P_{p},
\end{align}
where $V$ represents the velocity of the UAV, $P_{b}$, $P_{i}$, and $P_{p}$ are the blade profile power, induced power, and parasitic power, respectively. In particular, the $P_{b}$ and $P_{i}$ are modeled as
\begin{align}
    {P_{b}}=&{P_{0}}\left({1 + \frac {{3{V^{2}}}}{{\Omega ^{2}{r^{2}}}}}\right),  \\{P_{i}}=&{P_{s}}{\left({{\sqrt {1 + \frac {V^{4}}{4v_{0}^{4}}} - \frac {V^{2}}{2v_{0}^{2}}} }\right)^{1/2}},
\end{align}
where $P_{0}$ represents the hovering blade profile power, $P_{s}$ represents the hovering induced power, $v_{0}$ represents the induced velocity for rotor in forwarding flight, $\Omega$ and $r$ represent the blade angular velocity and the blade radius, respectively. In addition, the parasitic power is $P_{p} = \frac{1}{2}d_0\rho s_{p}AV^3$ with $s_{p}$ being the rotor solidity, $d_{0}$ being the fuselage drag ratio and $A$ and $\rho$ denote the rotor disc area and air density, respectively. Therefore, to ensure that the UAV can return to the UAV charging station before it runs out of energy, we have the following constraint
\begin{align}
    \frac{\Vert \mathbf{q}^{r}[n] \Vert}{V_{\max}}P(V_{\max}) + E_{cons}[n] \leq E_{\max},
\end{align}
where $E_{cons}[n]$ and $E_{\max}$ represent the consumed energy and the total energy of the UAV, respectively.

\begin{figure}[t]
	\centering
	\includegraphics[width=3.4in]{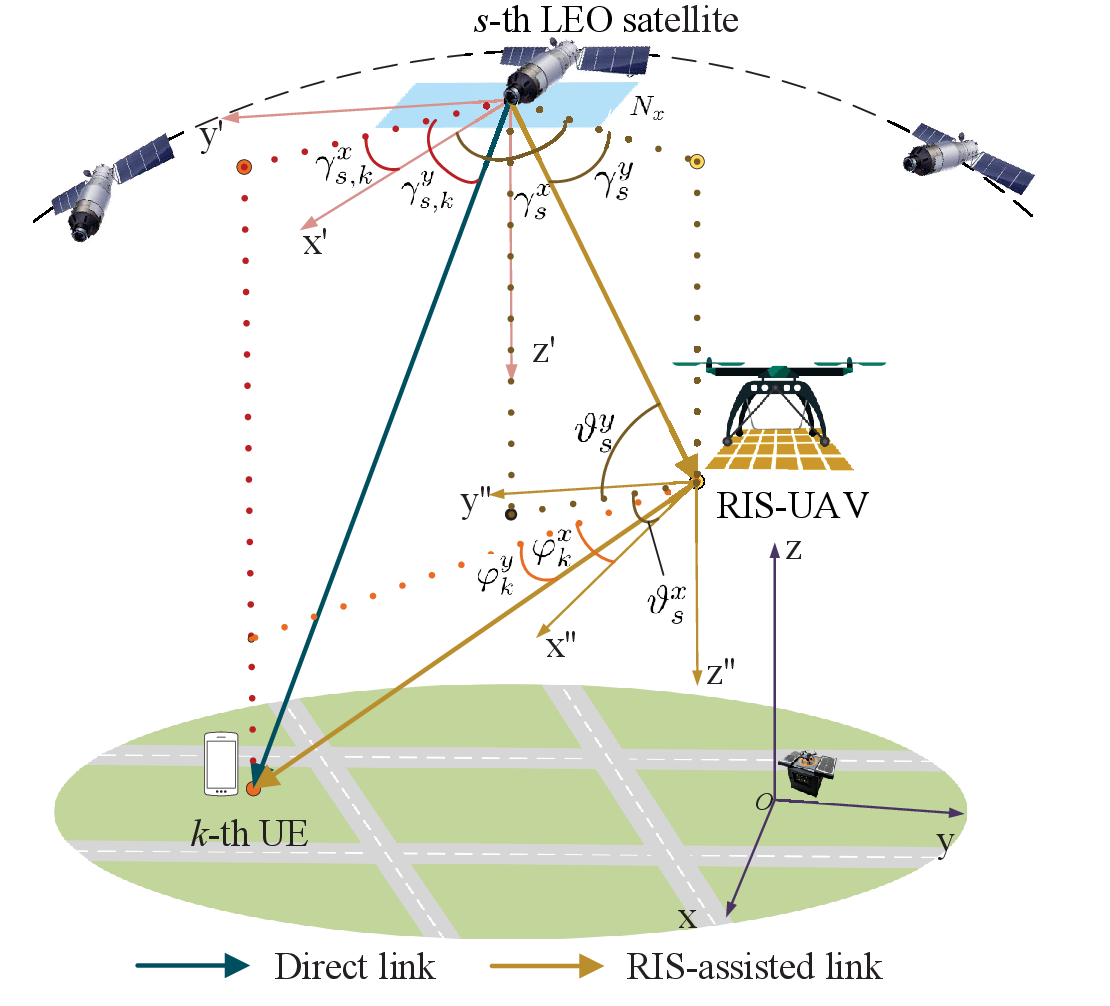}
	\caption{Angle relationship of the RIS-assisted satellite-terrestrial channel}
	\label{angle}
\end{figure}

In this context, the $s$-th LEO satellite constructs a transmitted signal $\mathbf{x}_s$ (omit time slot index [n] for brevity) as
\begin{align}\label{Transmit signal}
    \mathbf{x}_s = \sum_{k=1}^{K} \mathbf{v}_{s,k}e_{k},
\end{align}
where $\mathbf{v}_{s,k}$ represents the beamforming vector at the $s$-th satellite for transmitting the data symbol $e_{k}$ of the $k$-th UE. Herein, the data symbols $\{e_{k}\}_{k=1}^{K}$ satisfy $\mathbb{E}\{\vert e_{k}\vert^2\}=1$ and are statistically independent. Then, the signal is sent to UEs via RIS-assisted satellite-terrestrial channels, as shown in Fig. \ref{angle}.

For each RIS-assisted satellite-terrestrial channel, there are two links between the LEO satellite and the UE, i.e., a direct link and a RIS-assisted link. Over the direct link, the transmitted signal is usually subject to severe path-loss as well as weather effect. Therefore, the direct link between the $s$-th LEO satellite and the $k$-th UE can be formulated as \cite{Channel modeling 1}, \cite{Channel modelling 2}
\begin{align} \label{h}
    \mathbf{h}_{s,k} = L_{s,k}  \left(\sqrt{\frac{\kappa_{s,k}}{\kappa_{s,k}+1}}\mathbf{h}_{s,k}^{\text{LoS}} + \sqrt{\frac{1}{\kappa_{s,k}+1}}\mathbf{h}_{s,k}^{\text{NLoS}}\right),
\end{align}
where the $\kappa_{s,k}$ denotes the Rician factor, and $L_{s,k}$ denotes the large-scale fading component of the direct link, which is given by
\begin{align}
    L_{s,k} = \sqrt{\left(\frac{\lambda}{4\pi d_{s,k}}\right)^2 G_{k}} r_{s,k} \sqrt{b_{s,k}},
\end{align}
where $\lambda$ denotes the signal wave-length, $d_{s,k}$ denotes the distance between the $s$-th LEO satellite and the $k$-th UE, $G_k$ denotes the receive antenna gain at the $k$-th UE. In addition, $r_{s,k} = \xi_{s,k}^{1/2} e^{-j\varrho_{s,k}}$ denotes the rain attenuation with $\varrho_{s,k}$ being the phase vector and $\xi_{s,k}$ being the rain attenuation power gain, which is of the form of dB following the lognormal random distribution, i.e., $\ln(\xi^{\text{dB}}_{s,k})\sim \mathcal{N}( \mu_r, \sigma_r^2)$. Moreover, $b_{s,k}$ is the satellite antenna gain, which is given by \cite{channel yingming}
\begin{align}
    b_{s,k} = b_{s}^{\max}\left(\frac{J_1(u_{s,k})}{2u_{s,k}}+36\frac{J_3(u_{s,k})}{u_{s,k}} \right)^3,
\end{align}
where $b_{s}^{\max}$ denotes the maximum antenna gain, $u_{s,k} = 2.071(\sin(\varepsilon_{s,k})/\sin(\varepsilon^{\text{3dB}}_{s}))$ with $\varepsilon_{s,k}$ and $\varepsilon^{\text{3dB}}_{s}$ denoting the off-axis angle of the $k$-th UE for the beam boresight and the 3-dB angle of the $s$-th satellite, respectively. Moreover, $\mathbf{h}^{\text{LoS}}_{s,k}$ and $\mathbf{h}^{\text{NLoS}}_{s,k}$ denote the LoS and non-line-of-sight (NLoS) components of the direct link, respectively. The NLoS component $\mathbf{h}_{s,k}^{\text{NLoS}}$ follows a circularly symmetric complex Gaussian distribution $\mathcal{CN}(0, \mathbf{J}_{N_t \times 1})$ \cite{Channel modeling 1}. The LoS component $\mathbf{h}_{s,k}^{\text{LoS}}$ is equal to the UPA response of the LEO satellite, which is given by \cite{Channel modelling 3}
\begin{align}
    \mathbf{h}^{\text{LoS}}_{s,k} = \mathbf{a}^s_{x}(\gamma^x_{s,k},\gamma^y_{s,k}) \otimes \mathbf{a}^s_y(\gamma^y_{s,k}),
\end{align}
where $\gamma^x_{s,k}$ and $\gamma^y_{s,k}$ are the azimuth and elevation angles of the transmitted signal, respectively. Moreover, $\mathbf{a}^s_{x}(\gamma^x_{s,k},\gamma^y_{s,k})$ and $\mathbf{a}^s_y(\gamma^y_{s,k})$ are the array responses of $x'$-axis and $y'$-axis of the $s$-th LEO satellite UPA, given by
\begin{align}
    \mathbf{a}^s_{x}(\gamma^x_{s,k},\gamma^y_{s,k}) = \sqrt{\frac{1}{N_x}} [1; e^{-j \frac{2\pi}{\lambda} d_x \sin(\gamma^x_{s,k})\cos(\gamma_{s,k}^y)};\notag \\ \cdots;
    e^{-j \frac{2\pi}{\lambda} d_x  (N_x - 1)\sin(\gamma^x_{s,k})\cos(\gamma_{s,k}^y)}],
\end{align}
and
\begin{align}
    \mathbf{a}^s_y(\gamma^y_{s,k}) = \sqrt{\frac{1}{N_y}}[1; e^{-j \frac{2\pi}{\lambda} d_y  \cos(\gamma_{s,k}^y)}; \notag \\\cdots;e^{-j \frac{2\pi}{\lambda} d_y (N_y - 1) \cos(\gamma_{s,k}^y)}],
\end{align}
where the $N_x$ and $N_y$ are the number of columns and rows of the LEO satellite UPA, respectively, and $d_x$ and $d_y$ are the antenna spacing on the $x'$-axis and $y'$-axis of the LEO satellite UPA, respectively. In particular, the $x'$-axis and $y'$-axis are the column and row direction of the UPA of LEO satellite, respectively. Notice that the row direction $y'$-axis is parallel to the motion direction of LEO satellite.

For a RIS-assisted link, it includes a sub-link from the LEO satellite to the RIS-UAV and a sub-link from the RIS-UAV to the UE. Similar to the direct link from the LEO satellite to UE, the sub-link from the $s$-th LEO satellite to the RIS-UAV can be formulated as
\begin{align}\label{G}
    \mathbf{G}_{s} = L^r_{s}  \left( \sqrt{\frac{\kappa^r_s}{\kappa^r_s+1}}\mathbf{G}^{\text{LoS}}_s + \sqrt{\frac{1}{\kappa^r_s+1}}\mathbf{G}_{s}^{\text{NLoS}} \right),
\end{align}
where $\kappa^r_s$ represents the Rician factor, and $L^r_s$ represents the large-scale fading component, which is given by
\begin{align}
    L^r_s = \frac{\lambda}{4\pi d_s^r} r_{s} \sqrt{b_s},
\end{align}
where $d_s^r$ is the distance between the $s$-th LEO satellite and the RIS-UAV, $r_s$ and $b_s$ are the rain attenuation and satellite antenna gain, respectively. Moreover, $\mathbf{G}_{s}^{\text{LoS}}\in \mathbb{C}^{M_r \times N_t}$ and $\mathbf{G}_{s}^{\text{NLoS}} \in \mathbb{C}^{M_r \times N_t} $ denote the LoS and NLoS components of the sub-link, respectively. Herein, the NLoS component $\mathbf{G}_{s}^{\text{NLoS}}$ follows the distribution of $\mathcal{CN}(0, \mathbf{J}_{M_r \times N_t})$. The LoS component $\mathbf{G}_{s}^{\text{LoS}}$ can be formulated as
\begin{align}
    \mathbf{G}_{s}^{\text{LoS}} = \mathbf{a}_{x}^r(\vartheta_{s}^{x},\vartheta_{s}^{y})\otimes \mathbf{a}^r_y(\gamma^y_{s,k}) \cdot (\mathbf{a}^s_{x}(\gamma^x_{s,k},\gamma^y_{s,k}) \otimes \mathbf{a}^s_y(\gamma^y_{s,k}))^{\text{H}},
\end{align}
where $\gamma^{x}_{s}$, $\gamma^{y}_{s}$, $\vartheta_{s}^{x}$, and $\vartheta_{s}^{y}$ are the azimuth and the elevation angles of departure and the azimuth and the elevation angles of arrive of the signal from the $s$-th satellite to the RIS-UAV, respectively. The RIS array responses $\mathbf{a}_{x}^r(\vartheta_{s}^{x},\vartheta_{s}^{y})$ ($x''$-axis) and $ \mathbf{a}^r_y(\vartheta_{s}^{y})$ ($y''$-axis) follow the same mathematical form as the satellite's array responses $\mathbf{a}^s_{x}(\gamma^x_{s,k},\gamma^y_{s,k}) $ and $ \mathbf{a}^s_y(\gamma^y_{s,k})$, respectively. In particular, the $x''$-axis and $y''$-axis are the row and column directions of the RIS elements, where the $x''$-axis has the same direction as the $x$-axis.

When the transmitted signals reach the RIS-UAV through the sub-link from the LEO satellite to the RIS-UAV, they will be transmitted by the RIS. Let $\theta_m$ denote the phase shift coefficient of the $m$-th element of the RIS, then the transmitting coefficient matrix of RIS can be expressed as \cite{RIS matrix}
\begin{align} \label{theta}
    \boldsymbol{\Theta} = \text{diag}  \{[e^{j\theta_1}, e^{j\theta_2},\cdots,e^{j\theta_{M_r}}]\}.
\end{align}
Finally, the signals transmitted by the RIS arrive at the UEs through the sub-link from the RIS-UAV to UEs. In particular, the sub-link from the RIS-UAV to the $k$-th UEs is formulated as \cite{Secure communication in UAV-RIS-empowered multiuser networks}
\begin{align}\label{g}
    \mathbf{g}_{k} = F_{k} \left(\sqrt{\frac{\nu_{k}}{\nu_{k}+1}}\mathbf{g}_{k}^{\text{LoS}}  + \sqrt{\frac{1}{\nu_{k}+1}}\mathbf{g}_{k}^{\text{NLoS}} \right),
\end{align}
where $F_k$ is the large-scale fading component, which is mainly determined by the propagation distance. Thus, $F_k$ is formulated as
\begin{align}
    F_k = \sqrt{\left(\frac{\lambda}{4\pi \Vert \mathbf{q}^{r} - \mathbf{q}_k^{u} \Vert}\right)^2G_{k}}.
\end{align}
In addition, $\nu_{k}$ denotes the Rician factor, $\mathbf{g}_{k}^{\text{NLoS}}$ and $\mathbf{g}_{k}^{\text{LoS}}$ represent the LoS and NLoS components, respectively. Similarly, $\mathbf{g}_{k}^{\text{NLoS}}$ follows a circularly symmetric complex Gaussian distribution $\mathcal{CN}(0, \mathbf{J}_{M_r\times 1})$, and $\mathbf{g}_{k}^{\text{LoS}}$ is equivalent to the UPA response of RIS, which is given by
\begin{align}
    \mathbf{g}_{k}^{\text{LoS}} = \mathbf{a}_{x}^r(\varphi_{s}^{x},\varphi_{s}^{y})\otimes \mathbf{a}^r_y(\varphi^y_{s,k}),
\end{align}
where $\varphi^{x}_{k}$ and $\varphi^{y}_{k}$ are the azimuth and elevation angles of the departure, respectively. All the angle relationships of the RIS-assisted satellite-terrestrial channel are summarised in Fig. \ref{angle}.

To sum up, the RIS-assisted satellite-terrestrial channel between the $s$-th LEO satellite and the $k$-th UE can be equivalent to\footnote{Because of the low flight altitude and low flight speed of the RIS-UAV, the UAV's motion induces only a minimal additional doppler shift and can be treated as a part of the Doppler spread effect. Therefore, we assume that the directed link has the same Doppler shift and propagation delay as the RIS-assisted link.}
\begin{align}
    \mathbf{f}_{s,k} = \mathbf{h}_{s,k} + \mathbf{G}_{s}^{\text{H}}\boldsymbol{\Theta}^{\text{H}}\mathbf{g}_k.
\end{align}
In this context, the received signal at the $k$-th UE can be expressed as
\begin{align}
    y_{k}(t) =& \sum_{s=1}^{S}\mathbf{f}_{s,k}^{\text{H}} \mathbf{x}_s(t-\iota_{s,k})e^{j2\pi f_{s,k}^\text{D}t} + n_{k}\notag  \\
    = &\underbrace{\sum_{s=1}^{S}\mathbf{f}^{\text{H}}_{s,k}\mathbf{v}_{s,k}e_{k}(t-\iota_{s,k})e^{j2\pi f_{s,k}^\text{D}t}}_{\text{desired signal}} \\&+ \underbrace{\sum_{s=1}^{S}\mathbf{f}^{\text{H}}_{s,k}\sum_{l=1,l\neq k}^{K}\mathbf{v}_{s,l}e_{l}(t-\iota_{s,k})e^{j2\pi f_{s,k}^\text{D}t}}_{\text{interference}} \notag \\
    &+ \underbrace{n_{k}}_{\text{noise}},
\end{align}
where $n_{k}$ is additive white Gaussian noise (AWGN) with zero mean and variance $\sigma_{k}^2$. In particular, $\sigma_{k}^2 = \kappa B\zeta$ with $\kappa$, $B$, and $\zeta$ denoting the Boltzmann constant, channel bandwidth and noise temperature, respectively. $f_{s,k}^\text{D} $ and $\iota_{s,k}$  denote the Doppler shift and propagation delay between the transmission from the $s$-th satellite to the $k$-th UE. The explicit expression of $f_{s,k}^{\text{D}}$ can be formulated as $f_{s,k}^{\text{D}} = f_{c}(v_{s}\cdot\sin{\gamma_{s,k}^{x}}\cdot\cos{\gamma_{s,k}^{y}}/c)$ with $v_{s}$ being the velocity of the $s$-th satellite and $f_{c}$ being the carrier frequency. The propagation delay $\iota_{s,k}$ depends on the relative signal transmission distance between the $s$-th satellite and the $k$-th UE. Since the distance between UEs is of the order of meters and is much smaller than the signal transmission distance, we assume that the Doppler shift and the propagation delay between different UEs are same for the same satellite, and different for different satellites, i.e. $f_{s}^{\text{D}} = f_{s,k}^{\text{D}},\forall k$ and $\iota_{s} = \iota_{s,k}, \forall k$.
The Doppler shift and propagation delay can be acquired by the LEO satellites through the CSI estimation process \cite{channel estimation}. In this case, joint time synchronization and Doppler compensation are pre-implemented at each satellite to ensure that all signals from the cooperative satellites can arrive at the UE simultaneously \cite{Time synchronization}, \cite{Doppler compensation}\footnote{The residual Doppler variation due to inter-UE geometric differences can be tolerated at the UE end, as it falls within the tracking capabilities of modern receivers, e.g., phase-locked loops or wideband synchronization algorithms.}. For ease of exposition, we omit the time idnex $t$ and define $\mathbf{v}_{k}  = [\mathbf{v}_{1,k} ;\cdots;\mathbf{v}_{S,k} ] \in \mathbb{C}^{N_tS\times 1}$ to indicate the LEO satellite group beamforming to the $k$-th UE. Similarly, the equivalent channel from the LEO satellite group to the $k$-th UE is expressed as $\mathbf{f}_{k} = [\mathbf{f}_{1,k};\cdots;\mathbf{f}_{S,k}] \in \mathbb{C}^{N_tS\times 1}$. As a result, the received signal at the $k$-th UE can be reformulated as
\begin{align}
y_{k}  =\mathbf{f}_{k}^{\text{H}} \mathbf{v}_{k}e_{k}  + \sum_{l\neq k}^{K} \mathbf{f}_{k}^{\text{H}} \mathbf{v}_{l}e_{l}  + n_{k}.
\end{align}
Under this circumstance, the ergodic rate for the $k$-th UE can be computed as
\begin{align}\label{ergodic data rate}
    R_{k} = \mathbb{E}\left\{\log_2\left(1+\frac{\vert\mathbf{f}_{k}^{\text{H}}\mathbf{v}_{k} \vert^2}{\sum\limits_{l\neq k}^{K}\vert\mathbf{f}^{\text{H}}_{k}\mathbf{v}_{l}\vert^2 + \sigma^2_{k}}\right)\right\}.
\end{align}

It is evident that the LEO satellite group beamforming $\mathbf{v}_{l}, \forall l$ and the equivalent channel from LEO satellite group to the $k$-th UE $\mathbf{f}_{k}$ are the critical factors determining the ergodic rate of the $k$-th UE. Consequently, it is anticipated that the communication quality of UEs can be enhanced by optimizing the LEO satellite beamforming, RIS phase shift, and RIS-UAV trajectory within the LEO satellite constellation communication system.

\section{Design of RIS-UAV-assisted LEO satellite constellation}
In this section, we focus on the optimization of the RIS-UAV-assisted LEO satellite constellation communication system. Specifically, we first formulate a UEs fairness communication problem of maximizing the minimum UE ergodic rate by optimizing LEO satellite beamforming, RIS phase shift, and RIS-UAV trajectory. Then, we propose an effective AO-based algorithm to solve the formulated optimization problem. Finally, we analyze the performance of the proposed algorithm.

\subsection{Problem Formulation}
To ensure UE fairness during a data frame period $T$, we aim to maximize the minimum ergodic rate among all UEs by optimizing LEO satellite beamforming, RIS phase shift, and RIS-UAV trajectory. Let $\mathbf{V} = \{\mathbf{v}_{k}\}_{k=1}^{K}$ and $\boldsymbol{\theta} = \{\boldsymbol{\theta}_{m}\}_{m=1}^{M_r}$ denote the set of LEO satellite group beamforming and RIS phase shift, respectively, the optimization problem can be formulated as the following max-min problem:
\begin{subequations}\label{OP0}
\begin{align}
    \text{(P1)}: \max_{\mathbf{V}, \boldsymbol{\theta}, \mathbf{q}^r}~&\min_{k} R_k \label{OP0_obj}\\
    \text{s.t.} ~& \text{(1), (2), (3), (7)}\label{OP0_con1}, \\
    & \text{tr}\left\{\sum_{k=1}^{K}\mathbf{v}_{s,k}\mathbf{v}_{s,k}^{\text{H}}\right\}\leq P_{s}^{m}, \forall s, \label{OP0_con2} \\
    & 0\leq\theta_{m} < 2\pi,\forall m,\label{OP0_con3}
\end{align}
\end{subequations}
where the objective function (\ref{OP0_obj}) stands for maximizing the minimum ergodic rate among all UEs, (\ref{OP0_con1}) denotes the motion limitation of the RIS-UAV, (\ref{OP0_con2}) denotes the transmit power budget constraint at LEO satellites with $P_{s}^m$ being the maximum transmit power of the $s$-th LEO satellite, and (\ref{OP0_con3}) denotes the phase shift constraint of RIS elements. It is seen that the objective function (\ref{OP0_obj}) is not a closed-form expression in terms of optimization variables due to the non-integrability of the internal function of the expectation operation, and the optimization variables $\mathbf{V}$, $\boldsymbol{\Theta}$, and $\mathbf{q}^r$ are coupled with each other in the expression (\ref{ergodic data rate}) for ergodic rate $R_{k}$, making it difficult to obtain an optimal solution to problem \text{(P1)} in polynomial time. To this end, we make some transformations on the optimization problem.

Firstly, to get a closed-form expression for ergodic rate of UE, we approximate the ergodic rate $R_{k}$ of the $k$-th UE as \cite{RIS statistical CSI}
\begin{align}\label{EDRA}
    R_{k} \approx \bar{R}_{k} = \log \left(1 + \frac{\mathbb{E}\left\{ \vert\mathbf{f}_{k}^{\text{H}}\mathbf{v}_{k} \vert^2 \right\}}{\mathbb{E}\left\{ \sum\limits_{l\neq k}^{K}\vert\mathbf{f}^{\text{H}}_{k}\mathbf{v}_{l}\vert^2 \right\} + \sigma^2_{k}}\right).
\end{align}
By substituting the direct link and RIS-assisted link equations (\ref{h}), (\ref{G}), and (\ref{g}) into (\ref{EDRA}), the approximated ergodic rate of the $k$-th UE can be further represented as
\begin{align}
    \bar{R}_{k} = \log_2\left(1 + \bar{\Gamma}_{k}\right),
\end{align}
where the expression for $\bar{\Gamma}_{k}$ is given at the top of the next page. In particular, the derivation of $\bar{\Gamma}_{k}$ and the definition of auxiliary variables are detailed in Appendix A.
\begin{figure*}[!t]
	\begin{align}\label{SINR expression}
        \bar{\Gamma}_{k} = \frac{\vert (\bar{\mathbf{h}}_{k}^{\text{H}} + \bar{\mathbf{g}}_{k}^{\text{H}}\boldsymbol{\Theta}\bar{\mathbf{G}})\mathbf{v}_{k} \vert^2 + \frac{1}{\nu_k+1}P^2_k \Vert \bar{\mathbf{G}}\mathbf{v}_k \Vert^2 +  \Vert\mathbf{a}_{k} \odot \mathbf{v}_{k}\Vert^2 + \Vert\mathbf{b}_{k}\odot\mathbf{v}_{k} \Vert^2}{\sum\limits_{l \neq k}^{K} \left(\vert (\bar{\mathbf{h}}_{k}^{\text{H}} + \bar{\mathbf{g}}_{k}^{\text{H}}\boldsymbol{\Theta}\bar{\mathbf{G}})\mathbf{v}_{l} \vert^2 + \frac{1}{\nu_k+1}P^2_k \Vert \bar{\mathbf{G}}\mathbf{v}_l \Vert^2 + \Vert\mathbf{a}_{k} \odot \mathbf{v}_{l}\Vert^2 + \Vert\mathbf{b}_{k}\odot\mathbf{v}_{l} \Vert^2 \right) + \sigma_k^2}.
    \end{align}
	{\noindent} \rule[-10pt]{18cm}{0.05em}
\end{figure*}
It is seen that the approximated ergodic rate of UE $\bar{R}_{k}$ has a closed-form expression in terms of the optimization variables. Thus, by introducing an auxiliary variable $t$, the original problem \text{(P1)} can be transformed as
\begin{subequations}\label{OP0.1}
    \begin{align}
        \text{(P1.1)}:\max_{\mathbf{V}, \boldsymbol{\Theta},\mathbf{q}^r} &t\\
        \text{s.t.} ~& (\ref{OP0_con1}),(\ref{OP0_con2}),(\ref{OP0_con3}), \\
        ~&  t \leq \bar{R}_k, \forall k. \label{OP0.1_con3}
    \end{align}
\end{subequations}
Since the problem \text{(P1.1)} is not jointly convex with respect to the three optimization variables, we decompose it into three sub-problems by using the AO method, i.e., LEO satellite beamforming design, RIS phase shift design, and RIS-UAV trajectory design. These sub-problems are iteratively optimized until convergence.

\subsection{LEO Satellite Beamforming Design}
Firstly, consider the sub-problem of the LEO satellite beamforming $\mathbf{V}$ design with fixed RIS phase shift $\boldsymbol{\theta}$ and RIS-UAV position $\mathbf{q}^r$, which can be reformulated as
\begin{subequations}\label{OP1.0}
    \begin{align}
        \text{(P2)}:\max_{\mathbf{V}} ~&t\\
        \text{s.t.} ~& \text{(\ref{OP0_con2})}, \text{(\ref{OP0.1_con3})},
    \end{align}
\end{subequations}
which is a non-convex problem. To this end, we adopt the semi-definite relaxation (SDR) technique to tackle the optimization problem \text{(P2)} \cite{Joint Communication Beamforming and Sensing Waveform Design}. By defining $\mathbf{V}_{k} = \mathbf{v}_{k}\mathbf{v}_{k}^{\text{H}}$ satisfying $\text{Rank}(\mathbf{V_{k}}) = 1$ and $\mathbf{V}_{k} \succeq 0$, the LEO satellite transmit power budget constraint (\ref{OP0_con2}) can be re-expressed as
\begin{align} \label{OP1_con2.1}
    \text{tr} \left\{ \boldsymbol{\Lambda}_s \sum_{k=1}^{K} \mathbf{V}_{k} \boldsymbol{\Lambda}_s \right\} \leq P_s^{\text{m}}, \forall s,
\end{align}
where $\Lambda_s = \boldsymbol{\tau}_{s} \otimes \mathbf{I}_{N_t} \in \mathbb{C}^{N_t\times SN_t}$ with $\boldsymbol{\tau}_{s} \in \mathbb{C}^{1 \times S}$ whose $s$-th element is 1 and rest are 0. Then, by introducing auxiliary variables $\mathbf{F}_{k} = (\bar{\mathbf{h}}_{k}^{\text{H}} + \bar{\mathbf{g}}_{k}^{\text{H}}\boldsymbol{\Theta}\bar{\mathbf{G}})(\bar{\mathbf{h}}_{k}^{\text{H}} + \bar{\mathbf{g}}_{k}^{\text{H}}\boldsymbol{\Theta}\bar{\mathbf{G}})^{\text{H}}$, $\mathbf{Q} = \bar{\mathbf{G}}^{\text{H}}\bar{\mathbf{G}}$, $\mathbf{A}_{k} = \text{diag}^2\{\mathbf{a}_{k}\}$, and $\mathbf{B}_{k} = \text{diag}^2 \{\mathbf{b}_{k}\}$, the constraint (\ref{OP0.1_con3}) can be reformulated as (\ref{OP1_con1.1}) at the top of the next page.
\begin{figure*}[!t]
    \begin{align}\label{OP1_con1.1}
        &\frac{\text{tr}\{\mathbf{F}_{k}\mathbf{V}_{k}\} + \frac{1}{\nu_k+1}P^2_k \text{tr}\{\mathbf{Q}\mathbf{V}_{k}\} +  \text{tr}\{\mathbf{A}_{k}\mathbf{V}_{k}\} + \text{tr}\{\mathbf{B}_{k}\mathbf{V}_{k}\}}{2^t - 1} \geq \notag \\
        &\sum\limits_{l\neq k}^{K}\left\{\text{tr}\{\mathbf{F}_{k}\mathbf{V}_{l}\} + \frac{1}{\nu_k+1}P^2_k\text{tr}\{\mathbf{Q}\mathbf{V}_{l}\} +  \text{tr}\{\mathbf{A}_{k}\mathbf{V}_{l}\} + \text{tr}\{\mathbf{B}_{k}\mathbf{V}_{l}\} \right\} + \sigma^2_{k}, \forall k \in K.
    \end{align}
	{\noindent} \rule[-10pt]{18cm}{0.05em}
\end{figure*}
Unfortunately, the constraint (\ref{OP1_con1.1}) is still non-convex due to the coupled objective function $t$ and the optimization variable $\mathbf{V}_k$. Note that the constraint (\ref{OP1_con1.1}) is convex when $t$ is fixed. Moreover, the left-hand part of constraint (\ref{OP1_con1.1}) is monotonic with respect to $t$ when the LEO satellite beamforming is fixed. Therefore, the optimal value of objective function $t$ can be found via a bisection search, which is proved later in the convergence analysis subsection \cite{Max-Min fairness in IRS-aided multi-cell MISO systems}. Hence, the problem \text{(P2)} with the fixed $t$ is reformulated as
\begin{subequations}\label{OP1.1}
    \begin{align}
        \text{(P2.1)}:\text{Find}&~\{\mathbf{V}_{k}\}_{k=1}^{K} \label{OP1.1OBj}\\
        \text{s.t.}&~ (\ref{OP1_con2.1}), (\ref{OP1_con1.1}), \\
        &~ \text{Rank}(\mathbf{V}_{k})=1,\forall k, \label{OP1.1_con1}\\
        &~ \mathbf{V}_{k} \succeq 0, \forall k\label{OP1.1_con2}.
    \end{align}
\end{subequations}
However, the optimization problem \text{(P2.1)} remains non-convex owing to the rank-one constraint on the optimization variable $\mathbf{V}_{k}$ (\ref{OP1.1_con1}). To solve it, we adjust the objective function (\ref{OP1.1OBj}) to enforce the rank-one constraint, thus eliminating the non-convex constraint (\ref{OP1.1_con1}). It is known that $\mathbf{V}_k$ is a semi-positive definite matrix with all eigenvalues greater than or equal to zero. Thus, the rank-one constraint is equivalent to
\begin{align}\label{Rankone1}
    \text{tr}\{ \mathbf{V}_{k} \} - \lambda_{\max}\{ \mathbf{V}_{k} \} = 0,
\end{align}
where $\lambda_{\max}\{ \cdot \}$ denotes the operations of taking the maximum eigenvalue of the matrix. Since taking the maximum eigenvalue is not a smooth function, we utilize an iterative approximation form of the maximum eigenvalue to transform (\ref{Rankone1}) into a convex function, namely
\begin{align}
    \text{tr}\{\mathbf{V}^{j+1}_{k}\} - \left(\mathbf{v}^{j}_{k}\right)^{\text{H}}\mathbf{V}_{k}\mathbf{v}_{k}^{j} \geq \text{tr}\{\mathbf{V}^{j+1}_{k}\} - \lambda_{\max}\{ \mathbf{V}^{j+1}_{k} \} \geq 0,
\end{align}
where $\mathbf{v}_{k}^{j}$ represents the unit eigenvector corresponding to the largest eigenvalue of $\mathbf{V}^{j}_{k}$ obtained after the $j$-th iteration. In order to satisfy the rank-one constraints, it is necessary to make $\text{tr}\{\mathbf{V}^{j+1}_{k}\} - \left(\mathbf{v}^{j}_{k}\right)^{\text{H}}\mathbf{V}^{j+1}_{k}\mathbf{v}_{k}^{j} $ close to 0 as possible. Eventually, the problem with the fixed $t$ is transformed as
\begin{subequations}\label{OP1.2}
    \begin{align}
        \text{(P2.2)}: \min_{\mathbf{V}_{k}}~~ & \sum_{k=1}^{K}\text{tr}\{\mathbf{V}^{j+1}_{k}\} - \left(\mathbf{v}^{j}_{k}\right)^{\text{H}}\mathbf{V}^{j+1}_{k}\mathbf{v}_{k}^{j} \label{P1.3Obj}\\
     \text{s.t.}~~ &(\ref{OP1_con2.1}), (\ref{OP1_con1.1}),  (\ref{OP1.1_con2}),
    \end{align}
\end{subequations}
which is a convex problem and can be readily addressed with a mathematical toolbox, such as CVX. For the requirement of minimum approximated UE ergodic rate reaches $t$, if problem \text{(P2.2)} can be iteratively solved with the objective function converges to 0, then the LEO satellite beamforming solution $\mathbf{v}_{k}$ can be obtained by utilizing eigenvalue decomposition (EVD) of problem \text{(P2)} optimal solution $\mathbf{V}^*_{k}$, i.e. $\mathbf{v}_{k} = \sqrt{\lambda_{\max}\{ \mathbf{V}^*_{k} \}}\mathbf{v}^*_{k}$. Eventually, we can obtain the optimal LEO satellite beamforming solution to the problem \text{(P2)} via the bisection search method within the search range $(t_{\min}, t_{\max})$ until it reaches the predefined accuracy $\delta$, which is summarized in Algorithm 1. In particular, the upper bound $t_{\max}$ is theoretically derived as
\begin{align}
    t_{\max} = \min\limits_{k}\{\log_2\left(1 +  \frac{P_{s}\lambda_{\max}(\boldsymbol{\digamma}_{k})}{\sigma_{k}^2}\right)\},
\end{align}
where $\boldsymbol{\digamma}_{k} = \mathbf{F}_{k} + \frac{1}{\nu_k+1}P^2_k\mathbf{Q} + \mathbf{A}_{k} + \mathbf{B}_{k}$.

\begin{algorithm}
    \renewcommand{\algorithmicrequire}{\textbf{Input:}}
    \renewcommand{\algorithmicensure}{\textbf{Output:}}
    \caption{LEO satellite beamforming design}
    \label{alg:1}
    \begin{algorithmic}[1]
        \REQUIRE $\mathbf{h}_{s,k}$, $\mathbf{g}_{k}$, $\mathbf{G}_{s}, N_t, N_s, M_r, K, \sigma_{k}^2$, $\mathbf{q}^r$, $\boldsymbol{\theta}$, $t_{\min}$ ,$t_{\max}$;
	\ENSURE $\mathbf{V}$;
        \REPEAT
            \STATE $t = (t_{\max} + t_{\min}) / 2$;
            \STATE Initilize iteration index $j=0$, initial feasible points $\mathbf{v}^{0}_{k}$;
            \REPEAT
                \STATE solve problem \text{(P2.2)} for $\mathbf{V}_{k}^{j+1}$;
                \IF{{ problem \text{(P2.2)}} solved}
                    \STATE Update $\mathbf{v}_{k}^{j+1}$ by EVD of $\mathbf{V}_{k}^{j+1}$;
                \ELSE
                    \STATE break;
                \ENDIF
                \STATE update $j=j+1$
            \UNTIL{the objective value of the { problem \text{(P2.2)}} is convergent.}
            \IF{tr$\{\mathbf{V}^{*}_{k}\} - \lambda_{\max}\{\mathbf{V}^{*}_{k}\} \approx 0$}
                \STATE Update $\mathbf{v}_{k}$ by EVD of $\mathbf{V}_{k}^{*}$;
                \STATE $t_{\min} = t$;
            \ELSE
                \STATE $t_{\max} = t$;
            \ENDIF
        \UNTIL{$t_{\text{max}} - t_{\text{min}} \leq \delta$};
    \STATE \text{OUTPUT:} $\mathbf{V}$.
    \end{algorithmic}
\end{algorithm}

\subsection{RIS Phase Shift Design}
For any given beamforming $\mathbf{V}$ and RIS-UAV position $\mathbf{q}^{r}$, the RIS phase shift optimization sub-problem can be expressed as
\begin{subequations}
    \begin{align}\label{OP2}
        \text{(P3)} : \max_{\boldsymbol{\theta}} &~ t\\
        \text{s.t.} &~ (\ref{OP0_con3}), (\ref{OP0.1_con3}).
    \end{align}
\end{subequations}
Note that the problem \text{(P3)} has a similar structure with the problem \text{(P2)}, which means that the problem \text{(P3)} can be also tackled by the bisection search method over $t$. In this way, the RIS phase shift optimization sub-problem with a fixed $t$ is reformulated as
\begin{subequations}\label{OP2.2}
    \begin{align}
        \text{(P3.1)}:\text{Find}~&\{\boldsymbol{\theta}\}\\
        \text{s.t.}&~ (\ref{OP0_con3}), (\ref{OP0.1_con3}),
    \end{align}
\end{subequations}
which remains non-convex because the constraint (\ref{OP0.1_con3}) is non-convex in terms of phase shift $\theta_{m}$. Considering the difficulty in optimizing the phase shift coefficient $\theta_{m}$, which is an exponent of $e$, we define $\boldsymbol{\varphi} = \text{vec}(\boldsymbol{\Theta}) \in \mathbb{C}^{1\times M_r}$ to represent the transmitting coefficient vector of RIS. Consequently, the constraint (\ref{OP0_con3}) can be rewritten as
\begin{align} \label{RIS_transmitting_cofficient}
    \vert\boldsymbol{\varphi}_{1,m}\vert = 1.
\end{align}
In addition, by introducing $\boldsymbol{\Phi}_{k} = \text{diag}\{\bar{\mathbf{g}}_{k}^{\text{H}}\}\bar{\mathbf{G}}$, $\mathbf{b}_{k,l} = \boldsymbol{\Phi}_{k}\mathbf{v}_{l}$ and $q_{k,l} = \bar{\mathbf{h}}^{\text{H}}_{k}\mathbf{v}_{l}$, constraint (\ref{OP0.1_con3}) can be represented as
\begin{align}\label{SINR2}
    2^t-1 \leq \frac{\boldsymbol{\varphi} \mathbf{B}_{k,k}\boldsymbol{\varphi}^{\text{H}} + 2\text{Re}\{\boldsymbol{\varphi}\mathbf{p}_{k,k}\} + \Psi_{k,k}}{\sum\limits_{l\neq k}^{K} \boldsymbol{\varphi} \mathbf{B}_{k.l}\boldsymbol{\varphi}^{\text{H}} + 2\text{Re}\{\boldsymbol{\varphi}\mathbf{p}_{k,l}\} + \Psi_{k,l} + \sigma_{k}^2},
\end{align}
where $\mathbf{B}_{k,l} = \mathbf{b}_{k,l}\mathbf{b}_{k,l}^{\text{H}}$, $\mathbf{p}_{k,l} = \mathbf{b}_{k,l} q^{\text{H}}_{k,l}$ and $ \Psi_{k,l} = \vert q_{k,l}\vert^2 + \frac{1}{\nu_k+1}P^2_k \Vert \bar{\mathbf{G}}\mathbf{v}_l \Vert^2 + \Vert\mathbf{a}_{k}\odot \mathbf{v}_{l}\Vert^2 + \Vert\mathbf{b}_{k}\odot \mathbf{v}_{l} \Vert^2$. Note that the numerator and denominator of the fractional part of expression (\ref{SINR2}) contain both quadratic and primary terms of the $\boldsymbol{\varphi}$. Inspired by \cite{RISSDR}, we introduce  the auxiliary variable $\bar{\boldsymbol{\varphi}} = [\boldsymbol{\varphi},1]$, and define $\boldsymbol{\phi} = \bar{\boldsymbol{\varphi}}^{\text{H}}\bar{\boldsymbol{\varphi}}$ satisfying $\text{Rank}\{\boldsymbol{\phi}\} = 1$, $\boldsymbol{\phi}_{m,m}=1$ and $\boldsymbol{\phi} \geq 0 $ to rewrite the constraint (\ref{SINR2}) as
\begin{align}\label{OP2.1con1.2}
    \frac{1}{2^t-1}(\text{tr}\{\boldsymbol{\phi}\boldsymbol{\Xi}_{k,k}\} + \Psi_{k,k}) - \sum\limits_{l\neq k}^{K}(\text{tr}\{\boldsymbol{\phi}\boldsymbol{\Xi}_{k,l}\} + \Psi_{k,l}) \geq
     \sigma_{k}^2,
\end{align}
where
\begin{align}
    \boldsymbol{\Xi}_{k,l} =
    \begin{bmatrix}
    \mathbf{B}_{k,l} & \mathbf{p}_{k,l}\\
    \mathbf{p}_{k.l}^{\text{H}} & 0
    \end{bmatrix}.
\end{align}
At this point, we have successfully transformed the non-convex constraint (\ref{OP0.1_con3}) into a convex form (\ref{OP2.1con1.2}). However, the rank-one constraint on $\mathbf{\boldsymbol{\phi}}$ still makes the problem non-convex. Similarly, we adjust the objective function to impose the rank-one constraint on $\mathbf{\boldsymbol{\phi}}$, as adopted in LEO satellite beamforming design. Thereby, the RIS phase shift optimization sub-problem with a fixed $t$ is rewritten as
\begin{subequations}
    \begin{align}
        \text{(P3.2)}: \min_{\boldsymbol{\phi}} ~~& \text{tr}\{\boldsymbol{\phi}^{j+1}\} - \bar{\boldsymbol{\varphi}}^{j} \boldsymbol{\phi}^{j+1}(\bar{\boldsymbol{\varphi}}^{j})^{\text{H}} \label{OP2.2obj}\\
        \text{s.t.}~&(\ref{OP2.1con1.2}), \\
        &\boldsymbol{\phi}_{m,m} = 1,\forall m,\\
        &\boldsymbol{\phi}\succeq 0,
    \end{align}
\end{subequations}
where $\bar{\boldsymbol{\varphi}}^{j}$ denotes the eigenvector corresponding to the largest eigenvalue of $\boldsymbol{\phi}^{j}$ obtained after the $j$-th iteration. For a given $t$, with the feasible solution $\boldsymbol{\phi}^*$ satisfying $\text{tr}\{\boldsymbol{\phi}^{*}\} - \lambda_{\max}\{\boldsymbol{\phi}^{*}\} \approx 0$ obtained from the iteration of solving problem \text{(P3.2)}, the RIS phase shift $\boldsymbol{\theta}$ can be recovered from it, which is given by
\begin{align}\label{recoverphaseshift}
    \boldsymbol{\theta} = \angle \bar{\boldsymbol{\varphi}}^*[1:M_r],
\end{align}
where $\bar{\boldsymbol{\varphi}}^*[1:M_r]$ means the first $M_r$ elements of the row vector $\bar{\boldsymbol{\varphi}}^*$. For the RIS phase shift optimization subproblem, the bisection upper bound is set as $t_{\max} = \min\limits_{k}\{\log_2\left(1 +  \frac{U_{k}}{J_{k} + \sigma_{k}^2}\right)\}$ with
\begin{align}
U_{k} &= (\vert\bar{\mathbf{h}}_{k}^{\text{H}}\mathbf{v}\vert + \sum\limits_{m=1}^{M_r}[\bar{\mathbf{g}}_{k}]_{n}[\bar{\mathbf{G}}\mathbf{v}]_{n})^2+   \frac{1}{\nu_k+1}P^2_k \Vert \bar{\mathbf{G}}\mathbf{v}_k \Vert^2 \notag \\ &+ \Vert\mathbf{a}_{k} \odot \mathbf{v}_{k}\Vert^2 + \Vert\mathbf{b}_{k}\odot\mathbf{v}_{k} \Vert^2,
\end{align}
and \begin{align}
J_{k} = \sum\limits_{l \neq k}^{K}\left(\frac{1}{\nu_k+1}P^2_k \Vert \bar{\mathbf{G}}\mathbf{v}_l \Vert^2 + \Vert\mathbf{a}_{k} \odot \mathbf{v}_{l}\Vert^2 + \Vert\mathbf{b}_{k}\odot\mathbf{v}_{l} \Vert^2\right).
\end{align}
Finally, the optimal RIS phase shift can be found by the bisection search method, which is summarised in Algorithm 2.
\begin{algorithm}
    \renewcommand{\algorithmicrequire}{\textbf{Input:}}
    \renewcommand{\algorithmicensure}{\textbf{Output:}}
    \caption{RIS phase shift design}
    \label{alg:2}
    \begin{algorithmic}[1]
          \REQUIRE $\mathbf{h}_{k}$, $\mathbf{g}_{k}$, $\mathbf{G}_{s}, N_t, N_s, M_r, K, \sigma_{k}^2$, $\mathbf{V}$, $\mathbf{q}^r$, $t_{\min}$, $t_{\max}$;
	\ENSURE $\boldsymbol{\theta}$;
        \REPEAT
            \STATE $t = (t_{\max} + t_{\min}) / 2$;
            \STATE Initilize $j = 0$, initial feasible point $\bar{\boldsymbol{\varphi}}^0$
            \REPEAT
                \STATE solve problem \text{(P3.2)} for $\boldsymbol{\phi}^{j+1}$;
                \IF{{problem \text{(P3.2)}} is solved}
                    \STATE Update $\bar{\boldsymbol{\varphi}^{j+1}}$ via the EVD of $\boldsymbol{\phi}^{j+1}$;
                \ELSE
                    \STATE break;
                \ENDIF
                \STATE Update $j = j + 1$
            \UNTIL{the objective value of the {problem \text{(P3.2)}} is convergence.}
            \IF{tr$\{\boldsymbol{\phi}^{*}\} - \lambda_{\max}\{\boldsymbol{\phi}^{*}\} \approx 0$}
                \STATE Update $\boldsymbol{\theta}$ with obtained $\bar{\boldsymbol{\varphi}}^*$ as (\ref{recoverphaseshift});
                \STATE $t_{\min} = t$;
            \ELSE
                \STATE $t_{\max} = t$;
            \ENDIF
        \UNTIL{$t_{\text{max}} - t_{\text{min}} \leq \delta$};
    \STATE \text{OUTPUT:} $\boldsymbol{\theta}$.
    \end{algorithmic}
\end{algorithm}

\subsection{RIS-UAV Trajectory Design}
With the LEO satellite beamforming $\mathbf{V}$ and RIS phase shift $\boldsymbol{\theta}$ obtained previously, in a manner similar to the formulation (\ref{OP2.2}), the RIS-UAV trajectory optimization sub-problem with the fixed $t$ can be reformulated as
\begin{subequations}
\begin{align}
    \text{(P4)}: \text{Find}& ~\mathbf{q}^r\\
    \text{s.t.} &~ \text{(1)-(3), (7)}, (\ref{OP0.1_con3}),
\end{align}
\end{subequations}
which is non-convex since the constraint (\ref{OP0.1_con3}) remains unchanged. Given that the expression of $\bar{\Gamma}_{k}$ (\ref{SINR expression}) is not a convex function in terms of the position of the RIS-UAV $\mathbf{q}^r$, it needs to be treated appropriately. Let $\mathbf{u}_{k}^{\text{H}} = \frac{\lambda}{4\pi}\sqrt{\frac{G_{k}\nu_{k}}{\nu_{k}+1}}\mathbf{g}^{\text{LoS}}_{k}\boldsymbol{\Theta}\bar{\mathbf{G}}$, $G'_{k} = \frac{\lambda^2 G_k}{(4\pi)^2(\nu_k+1)}$, $ \mathbf{r}_{k} = \left[\sqrt{\frac{1}{\kappa_{1,k}+1}}L_{1,k} \mathbf{J}_{N_t\times 1} ; \cdots ; \sqrt{\frac{1}{\kappa_{S,k}+1}}L_{S,k}\mathbf{J}_{N_t\times 1}\right]$ and $\mathbf{w}_{k} = (\frac{\lambda G_k}{4\pi})\left[\sqrt{\frac{M_r}{\kappa^r_{1} + 1}}L^r_1 \mathbf{J}_{N_t\times 1} ; \cdots ; \sqrt{\frac{M_r}{\kappa^r_{S} + 1}} L^r_S \mathbf{J}_{N_t\times 1}\right]$,
the $\bar{\Gamma}_{k}$ can be recasted as
\begin{align} \label{SINR4}
    \bar{\Gamma}_{k} = \frac{\Upsilon_{k,k} + \frac{1}{\Vert \mathbf{q}^r - \mathbf{q}^u_k\Vert}\chi_{k,k} +  \frac{1}{\Vert \mathbf{q}^r - \mathbf{q}^u_k\Vert^2} \psi_{k,k} }{\sum\limits_{l\neq k}^{K}\left( \Upsilon_{k,l} + \frac{1}{\Vert \mathbf{q}^r - \mathbf{q}^u_k\Vert}\chi_{k,l} + \frac{1}{\Vert \mathbf{q}^r - \mathbf{q}^u_k\Vert^2}\psi_{k,l}\right) + \sigma_k^2 },
\end{align}
where
\begin{align}
    \Upsilon_{k,l} &= \Vert \mathbf{r}_{k}\odot \mathbf{v}_{l}\Vert^2 + \vert \bar{\mathbf{h}}_{k}^{\text{H}}\mathbf{v}_{l}\vert^2, \\
    \chi_{k,l} &= 2\text{Re}\{\mathbf{u}_{k}\mathbf{v}_{l}\mathbf{v}_{l}^{\text{H}}\bar{\mathbf{h}}_{k}\},
\end{align}
and
\begin{align}
    \psi_{k,l} &= \vert\mathbf{u}_{k}\mathbf{v}_{l}\vert^2 + G'_{k}\Vert\bar{\mathbf{G}}\mathbf{v}_l \Vert^2  + \Vert\mathbf{w}_{k}\odot\mathbf{v}_{l} \Vert^2 .\label{Psi}
\end{align}
Following this, by substituting (\ref{SINR4}) into the constraint (\ref{OP0.1_con3}), we have the constraint (\ref{OP2.1_con2}) at the top of the next page.
\begin{figure*}[!t]
    \begin{align}\label{OP2.1_con2}
        \frac{1}{2^t - 1}(\Upsilon_{k,k} + \frac{1}{\Vert \mathbf{q}^r[n] - \mathbf{q}^u_k\Vert}\chi_{k,k} +  \frac{1}{\Vert \mathbf{q}^r[n] - \mathbf{q}_k\Vert^2} \psi_{k,k}) \geq \sum\limits_{l\neq k}^{K}\left( \Upsilon_{k,l} + \frac{1}{\Vert \mathbf{q}^r[n] - \mathbf{q}^u_k\Vert}\chi_{k,l} + \frac{1}{\Vert \mathbf{q}^r[n] - \mathbf{q}^u_k\Vert^2}\psi_{k,l}\right) + \sigma_k^2.
    \end{align}
	{\noindent} \rule[-10pt]{18cm}{0.05em}
\end{figure*}
Unfortunately, the constraint (\ref{OP2.1_con2}) remains non-convex owing to the presence of the fractional term of $\mathbf{q}^r$. To facilitate the problem, the slack variable $\beta_{k}$ and $c_{k}, \forall k$ are introduced to split the constraint (\ref{OP2.1_con2}) into the following constraints:
\begin{align}\label{OP3.1_con1}
    \beta_k \geq \sum\limits_{l\neq k}^{K}\left( \Upsilon_{k,l} + \chi_{k,l}c_k + \psi_{k,l}c_{k}^2 \right) + \sigma_k^2,
\end{align}
\begin{align}\label{OP3.1_con2}
    \Upsilon_{k,k} + \chi_{k,k}c_{k} + \psi_{k,k}c_k^2 \geq (2^t - 1)\beta_{k},
\end{align}
and
\begin{align}\label{OP3.1_con3}
    c_k \geq \frac{1}{\Vert \mathbf{q}^r - \mathbf{q^u_k} \Vert}.
\end{align}
It is worth noting that the constraint (\ref{OP3.1_con1}) is convex since the quadratic term coefficients $\psi_{k,l}$ of $c_k$ must be greater than zero, which can be easily verified from the formula (\ref{Psi}). Meanwhile, a first-order Taylor expansion method is adopted to deal with the non-convexity of the constraints (\ref{OP3.1_con2}) and (\ref{OP3.1_con3}). By expanding the variable $c_k$ at point $c_k^{\text{pre}} = \frac{1}{\Vert\mathbf{q}^{r}[n-1] - \mathbf{q}_{k}^\text{u} \Vert}$ for both constraints, we have
\begin{align}\label{OP3.1_con2.1}
    (2^t-1)\beta_{k} \leq
\chi_{k,k} + 2c_1^{\text{pre}}\psi_{k,k}(c_k-c_k^{\text{pre}}) \notag \\+ \Upsilon_{k,k} + c_k^{\text{pre}}\chi_{k,k} + (c_k^{\text{pre}})^2\psi_{k,k},
\end{align}
and
\begin{align}\label{OP3.1_con3.1}
    \Vert\mathbf{q}^r -\mathbf{q}^u_{k}\Vert^2 - 3(c_k^{\text{pre}})^{-2} + 2(c_k^{\text{pre}})^{-3}c_k \leq 0.
\end{align}
After the above transformation, the optimization problem \text{(P4)} can be rewritten as
\begin{subequations}
    \begin{align}
        \text{(P4.1)} :\text{Find:} & ~{\mathbf{q}^r,~\boldsymbol{\beta},~\mathbf{c}} \\
        \text{s.t.} &~ \text{(1)-(3), (7)}, (\ref{OP3.1_con1}), (\ref{OP3.1_con2.1}), (\ref{OP3.1_con3.1}).
    \end{align}
\end{subequations}
where $\boldsymbol{\beta}$ and $\mathbf{c}$ are the set of the auxiliary variables $\beta_{k}$ and $c_{k},\forall k$, respectively. Finally, the optimal RIS-UAV position can be obtained by solving the problem \text{(P4.1)} while searching $t$ until convergence, which is listed in Algorithm 3. In particular, the search upper bound is set as $t_{\max} = \min\limits_{k}\{\log_2\left(1 +  \frac{Y_{k}}{J_{k} + \sigma_{k}^2}\right)\}$, where $Y_{k} = (\vert\bar{\mathbf{h}}_{k}^{\text{H}}\mathbf{v}\vert + \vert\frac{V_{\max}}{\Vert\mathbf{q}^{r}-\mathbf{q}^{u}_{k}\Vert}\bar{\mathbf{g}_{k}}\boldsymbol{\Theta}\bar{\mathbf{G}}\vert)^2 + \frac{1}{\nu_k+1}P^2_k \Vert \bar{\mathbf{G}}\mathbf{v}_k \Vert^2 +  \Vert\mathbf{a}_{k} \odot \mathbf{v}_{k}\Vert^2 + \Vert\mathbf{b}_{k}\odot\mathbf{v}_{k} \Vert^2$.
\begin{algorithm}
    \renewcommand{\algorithmicrequire}{\textbf{Input:}}
    \renewcommand{\algorithmicensure}{\textbf{Output:}}
    \caption{RIS-UAV trajectory design}
    \label{alg:3}
    \begin{algorithmic}[1]
        \REQUIRE $\mathbf{h}_{k}$, $\mathbf{g}_{k}$, $\mathbf{G}_{s}, N_t, N_s, M_r, K, \sigma_{k}^2$, $\mathbf{v}_{k}$, $\boldsymbol{\theta}$,  $\mathbf{q}^r$, $t_{\min}$, $t_{\max}$;
	    \ENSURE $\mathbf{q}^r$;
        \REPEAT
            \STATE $(t = t_{\max} + t_{\min}) / 2$;
            \STATE Solve problem \text{(P4.1)} for $\mathbf{q}^r$;
            \IF{problem \text{(P4.1)} is solved}
                \STATE Update $\mathbf{q}^r$;
                \STATE $t_\text{min} = t$;
            \ELSE
                \STATE $t_\text{max} = t$;
            \ENDIF
        \UNTIL{$t_{\text{max}} - t_{\text{min}} \leq \delta$};
    \STATE \text{OUTPUT:} $\mathbf{q}^r$.
    \end{algorithmic}
\end{algorithm}

By iteratively solving the above three subproblems and updating variables $\mathbf{V}$, $\boldsymbol{\theta}$ and $\mathbf{q}^r$ until convergence, a feasible solution for LEO satellite beamforming, RIS phase shift and RIS-UAV trajectory can be obtained. The detailed steps of the AO-based algorithm are described in Algorithm 4.
\begin{algorithm}
    \renewcommand{\algorithmicrequire}{\textbf{Input:}}
    \renewcommand{\algorithmicensure}{\textbf{Output:}}
    \caption{AO-based joint design of satellite beamforming, RIS phase shift and UAV trajectory}
    \label{alg:4}
    \begin{algorithmic}[1]
        \REQUIRE $\mathbf{h}_{k}$, $\mathbf{g}_{k}$, $\mathbf{G}_{s}, N_t, N_s, M_r, K, \sigma_{k}^2$, $\mathbf{q}^r[0]$, $\mathbf{q}^{u}_{k}$;
	\ENSURE $\mathbf{q}^r$, $\boldsymbol{\theta}$, $\mathbf{V}$;
    \FOR{n = 1 : N}
        \STATE Update SCSI after the $(n-1)$-th time slot.
        \STATE Initilize the $t_{\min}$, $t_{\max}$, $l = 1$, $t^0=0$, initial feasible points $(\mathbf{q}^{r})^{0}$ and $\boldsymbol{\theta}^{0}$.
        \WHILE{$\delta_o > \Delta$ and $l \leq L$}
            \STATE Reset $t_{\max}$ and obtain $\mathbf{V}^{l}$ according to Algorithm 1 with given $(\mathbf{q}^{r})^{l-1}$, $\boldsymbol{\theta}^{l-1}$, and $(t_{\min},t_{\max})$;
            \STATE Reset $t_{\max}$ and obtain $\boldsymbol{\theta}^{l}$ according to Algorithm 2 with given $(\mathbf{q}^{r})^{l-1}$ , $\mathbf{V}^{l}$, and $(t_{\min},t_{\max})$;
            \STATE Reset $t_{\max}$ and obtain $(\mathbf{q}^{r})^{l}$ according to Algorithm 3 with given $\boldsymbol{\theta}^{l}$ ,  $\mathbf{V}^{l}$, and $(t_{\min},t_{\max})$;
            \STATE Update $t^l = t_{\min}$, $\delta_o = t^{l} - t^{l-1}$ and $l=l+1$;
        \ENDWHILE
    \ENDFOR
    \end{algorithmic}
\end{algorithm}

\subsection{Algorithm Analysis}
\subsubsection{Convergence Analysis}
Firstly, we prove the convergence of the Algorithms 1-3. Since the three algorithms have the same way to get the optimal $t$, i.e., the bisection search method for $t$, we take Algorithm 1 as an example. In Algorithm 1, the constraint (\ref{OP1_con1.1}) is the only constraint that contains the objective function $t$ and is monotonic to $t$. Suppose we obtain a solution $\mathbf{V}^l$ by solving the problem \text{(P2.2)} until convergence at any given $t = t_1$. Because of the monotonicity of constraint (\ref{OP1_con1.1}) with respect to $t$, $\mathbf{V}^l$ is a feasible solution to problem \text{(P2.2)} for any $t < t_1$. Conversely, if there is not a feasible solution to the problem \text{(P2.2)} at $t = t_1$, not a feasible solution to the problem \text{(P2.2)} can be found for any $t > t_1$ consequently. Therefore, we can make $t$ converge by a bisection search method and finally get the nearly optimal $t$ in all three algorithms. Next, we prove the convergence of Algorithm 4, which calls Algorithms 1-3 to obtain feasible solutions for the three optimization variables $\mathbf{V}$, $\boldsymbol{\theta}$, and $(\mathbf{q}^{r})$. Let $f(\mathbf{V}^{l}, \boldsymbol{\theta}^{l}, (\mathbf{q}^{r})^{l})$ denote the minimum UE approximated ergodic rate with the obtained $\mathbf{V}^{l}$, $\boldsymbol{\theta}^{l}$, $(\mathbf{q}^{r})^{l}$ after the $l$-th iteration. According to the update steps in Algorithm 4, the feasible solution $\mathbf{V}^{l}$ can be obtained by Algorithm 1 with given $\mathbf{q}^{l-1}$, $\boldsymbol{\theta}^{l-1}$ after searching $t$ in the range $(t_{\min},t_{\max})$. Since the $t_{\min}$ is the max-min UE approximated ergodic rate with $\mathbf{V}^{l-1}$, $\boldsymbol{\theta}^{l-1}$, $(\mathbf{q}^{r})^{l-1}$ and $t_{\max}$ has been resettled to the theoretical maximum rate, we have the following inequality
\begin{align}\label{converge1}
 f(\mathbf{V}^{l}, \boldsymbol{\theta}^{l-1}, (\mathbf{q}^{r})^{l-1}) \geq f(\mathbf{V}^{l-1}, \boldsymbol{\theta}^{l-1}, (\mathbf{q}^{r})^{l-1}).
\end{align}
Similarly, after the steps 6 and 7, we have the following inequalities
\begin{align}\label{converge2}
 f(\mathbf{V}^{l}, \boldsymbol{\theta}^{l}, (\mathbf{q}^{r})^{l-1}) \geq f(\mathbf{V}^{l}, \boldsymbol{\theta}^{l-1}, (\mathbf{q}^{r})^{l-1}),
\end{align}
and
\begin{align}\label{converge3}
 f(\mathbf{V}^{l}, \boldsymbol{\theta}^{l}, (\mathbf{q}^{r})^{l}) \geq f(\mathbf{V}^{l}, \boldsymbol{\theta}^{l}, (\mathbf{q}^{r})^{l-1}).
\end{align}
Finally, with the inequalities (58), (59), and (60), we can get
\begin{align}
 f(\mathbf{V}^{l}, \boldsymbol{\theta}^{l}, (\mathbf{q}^{r})^{l}) \geq f(\mathbf{V}^{l-1}, \boldsymbol{\theta}^{l-1}, (\mathbf{q}^{r})^{l-1}),
\end{align}
which means the minimum UE approximated ergodic rate is non-decreasing after each iteration. Crucially, this creates a monotonic improvement sequence where each iteration either maintains or enhances the objective value. In addition, the combination of bounded objective function due to finite transmit power, the closed feasible sets from continuous variable domains, and non-empty solution spaces ensured by proper constraint design, satisfies all conditions for convergence to a stationary point \cite{boundness}.

\subsubsection{Complexity Analysis}
Herein, we analyze the complexity of the proposed Algorithm 4. Since the steps of Algorithm 4 contain Algorithms 1-3, we first analyze the complexity of Algorithms 1-3, separately. The complexity of Algorithm 1 primarily arises from solving the problem \text{(P2.2)}, which is imposed by $K+S$ LMI constraints of size $N_tSK$ and $K$ LMI constraints of size $N_tS$. Based on the complexity analysis for the SDP in \cite{Complexity analysis}, the complexity of Algorithm 1 is $C_1 = \mathcal{O}(I_1 N_b N_t^2S^2(N_tSK+K^2)(K+S+1))$ where $I_1 = \sqrt{(N_tSK(K+S+1))}\ln(1/\varsigma)$ and $N_b$ denote the number of iterations in the bisection searching method. Similarly, the complexity of Algorithm 2 and Algorithm 3 can be derived as $C_2 = \mathcal{O}(I_2N_bM_r^2(K^3+K^2M_r+M_r+1))$ with $I_2 = \sqrt{3(M_r+1)}\ln(1/\varsigma)$ and $C_3 = \mathcal{O}(\sqrt{5K+3}\ln(1/\varsigma)N_b(K^2 + 10K))$, respectively. Furthermore, defining the number of iterations of the outer loop as $L_o$, the total complexity of Algorithm 4 can be computed as $\mathcal{O}(L_o(C1+C2+C3))$, which shows that the proposed Algorithm 4 has a polynomial time complexity. In addition, we have compared the computational complexity of the proposed method and the existing beamforming and UAV trajectory optimization approaches in Table \ref{Complexity_Comparison} at the top of the next page. It can be seen that our proposed method has the same or even lower computational complexity comparable to existing approaches while being highly scalable for large-scale RIS-assisted systems and long-time slots. Hence, it can be employed for the RIS-UAV-assisted LEO satellite constellation communication in practice.
\begin{table*}[ht]
    \caption{COMPLEXITY COMPARISON}
    \begin{center}
    \begin{tabular}{|c|c|c|}
    \hline
    Subproblem &  proposed method complexity & existing approach complexity\\
    \hline\hline
    (1): LEO satellite beamforming design & $\mathcal{O}(I_1 N_b N_t^2S^2(N_tSK+K^2)(K+S+1))$ & $\mathcal{O}((K(N_tS)^2+K(N_tS)^3))$ \cite{WMMSE}\\
    \hline
    (2): RIS phase shift design & $\mathcal{O}(I_2N_bM_r^2(K^3+K^2M_r+M_r+1))$ & $\mathcal{O}(M_r^4\ln(1/\varsigma))$ \cite{UAV-RIS-aided SATIN}\\
    \hline
    (3): UAV trajectory design & $\mathcal{O}(N\sqrt{5K+3}\ln(1/\varsigma)N_b(K^2 + 10K))$ & $\mathcal{O}((8N)^{3.5}\ln(1/\varsigma))$ \cite{Secure communication in UAV-RIS-empowered multiuser networks}\\
    \hline
    \end{tabular}
    \label{Complexity_Comparison}
    \end{center}
\end{table*}

\section{Simulation Result}
In this section, we provide experimental results to validate the effectiveness of proposed algorithms. In order to fulfill the requirements of global coverage as well as low latency communication, we consider the Walker Delta constellation configuration \cite{Walker Delta}. The detailed constellation parameters are listed in Table \ref{Table1}.
\begin{table}
    \centering
    \caption{PARAMETERS of LEO SATELLITE CONSTELLATION}
    \label{Table1}
    \begin{tabular}{|c|c|} \hline
    \textbf{Parameter} & \textbf{Value} \\ \hline
    Orbital altitude & 550km \\ \hline
    Number of orbital planes $N^{\#}$ & 36  \\ \hline
    Number of LEO satellites per orbital plane $P^{\#}$ & 22 \\ \hline
    Orbital inclination & $53^\circ$ \\ \hline
    Minimum elevation angle of UEs & $10^\circ$ \\ \hline
    Phase factor & 1 \\ \hline
    \end{tabular}
\end{table}
All LEO satellites use Ka-band for communications with a carrier frequency of 30 GHz. In this case, RIS with $M_r = 100$ elements has a size of 5cm $\times$ 5cm and can be mounted on the UAV \cite{Size of RIS}. The positions of all UEs follow a uniform distribution over the rectangular region [0,300]$\times$[0,300] and the position of the terrestrial region is set as $10^6 \times$ [-2.6610, 4.5050, -1.7249] of the Earth-Centered, Earth-Fixed (ECEF) coordinate system. The duration of a data frames is configured according to the visibility period of the LEO satellite group [53]. In particular, in order to adapt to the CSI variation due to the high-speed movement of the satellite and balance the computational complexity, we set the time slot duration as $\delta = 1$ s. Unless otherwise specified, the remaining simulation parameters are set as shown in Table \ref{Table2}. The simulations were performed using MATLAB R2022B on an AMD Ryzen 5 5600X CPU computer with 16GB RAM.
\begin{table}[!ht]
    \centering
    \caption{PARAMETER SETUP}
    \label{Table2}
    \begin{tabular}{|c|c|} \hline
    \textbf{Parameter} & \textbf{Value} \\ \hline
    Carrier frequency $f$ &	30 GHz  \\ \hline
    Spectrum bandwidth $B$ & 25 MHz  \\ \hline
    Number of LEO satellites in a group $S$ & 3 \\ \hline
    Number of UPA antennas $N_s$ & $4 \times 4$ \\ \hline
    Transmit power budget of LEO satellite $P_s^{m}$ & 40 dBm \\ \hline
    Number of UEs $K$ & 6\\ \hline
    Number of elements of RIS $M_r$ & 100 \\ \hline
    Transmit gain to noise temperature $G_k/T$ & 34 dB/K \\ \hline
    Satellite antenna gain $b_k$ & 20 dBi \\ \hline
    Boltzmann constant $\kappa$ & $1.38\times10^{-23} J/m$\\ \hline
    Rain attenuation mean $\mu_r$ & -2.6 dB \\ \hline
    Rain attenuation variance $\sigma_{r}^2$ & 1.63 dB \\ \hline
    3-dB angle $\chi_l$ & 0.4° \\ \hline
    Rician factor of $\mathbf{h}_{l,k}$ $\kappa_{l,k}$ & 30 \\ \hline
    Rician factor of $\mathbf{G}_{l}$ $\kappa^r_{l}$ & 30 \\ \hline
    Rician factor of $\mathbf{g}_{k}$ $\nu_{k}$ & 10 \\ \hline
    Maximum iteration number $T^{\max}$ & 10\\ \hline
    Flight radius of the RIS-UAV $l_{\max}$& 200 m\\\hline
    Duration of a data frame $T$ & 60 s \\\hline
    Number of time slots $N$ & 60 \\\hline
    Maximum speed of UAV $V_{\max}$ & 5 m/s \\\hline
    \end{tabular}
\end{table}

\begin{figure}[t]
	\centering
	\includegraphics[width=3.4in]{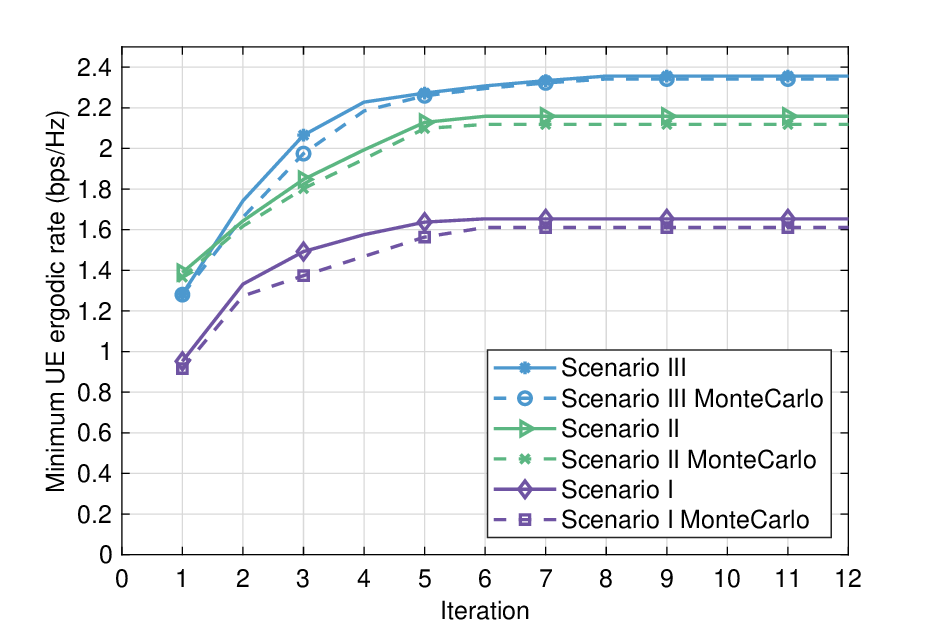}
	\caption{Convergence of proposed Algorithm 4.}
	\label{Convergence}
\end{figure}

\begin{figure}[t]
	\centering
	\includegraphics[width=3.4in]{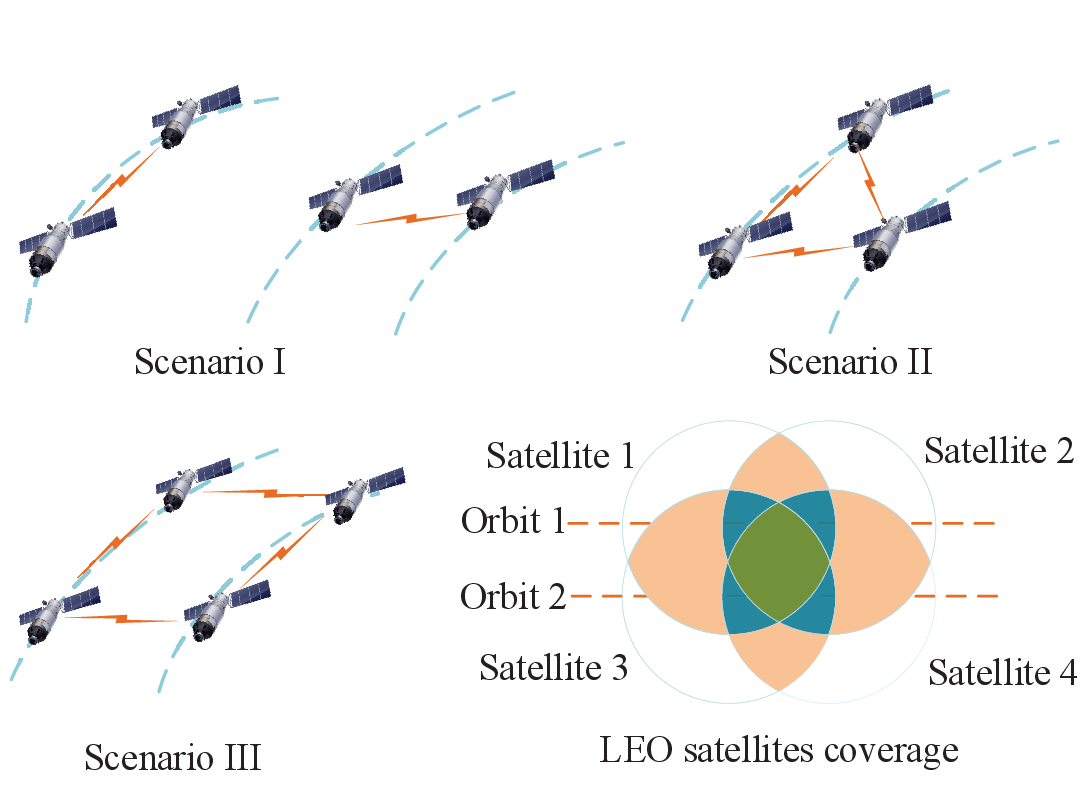}
	\caption{Types of LEO satellites collaboration.}
	\label{collaboration type}
\end{figure}
Firstly, we present the convergence behaviour of the Algorithm \ref{alg:4} with different LEO satellites collaboration scenarios in Fig. \ref{Convergence}. Without loss of generality, we take the first time slot as an example. In the considered Walker Delta satellite constellation configuration, when there are only two visible satellites, they are usually adjacent to each other in the same orbital plane or in adjacent orbits, which we refer to as Scenario I. When there are three visible satellites, they usually consist of two LEO satellites in the same orbit plus a satellite in an adjacent orbit, called Scenario II. When there are four visible satellites, they are usually uniformly distributed in two adjacent LEOs, called Scenario III. These LEO satellite collaboration scenarios and their corresponding coverage areas are shown in Fig. \ref{collaboration type}. It can be seen that the minimum UE ergodic rate increases monotonically with each iteration and eventually converges to a critical point within a few iterations for different numbers of LEO satellites. In addition, the minimum UE approximated ergodic rate increases as the number of visible satellites increases. The results confirm that the proposed Algorithm 4 has a fast convergence speed under different satellite collaboration scenarios. In order to testify the effectiveness of the proposed sCSI-based approach, we conduct the MonteCarlo simulation. The results demonstrate that the differences between the minimum UE approximated ergodic rate with their corresponding MonteCarlo results are negligible, confirming the effectiveness of the proposed sCSI-based approach.

\begin{figure}[t]
	\centering
	\includegraphics[width=3.4in]{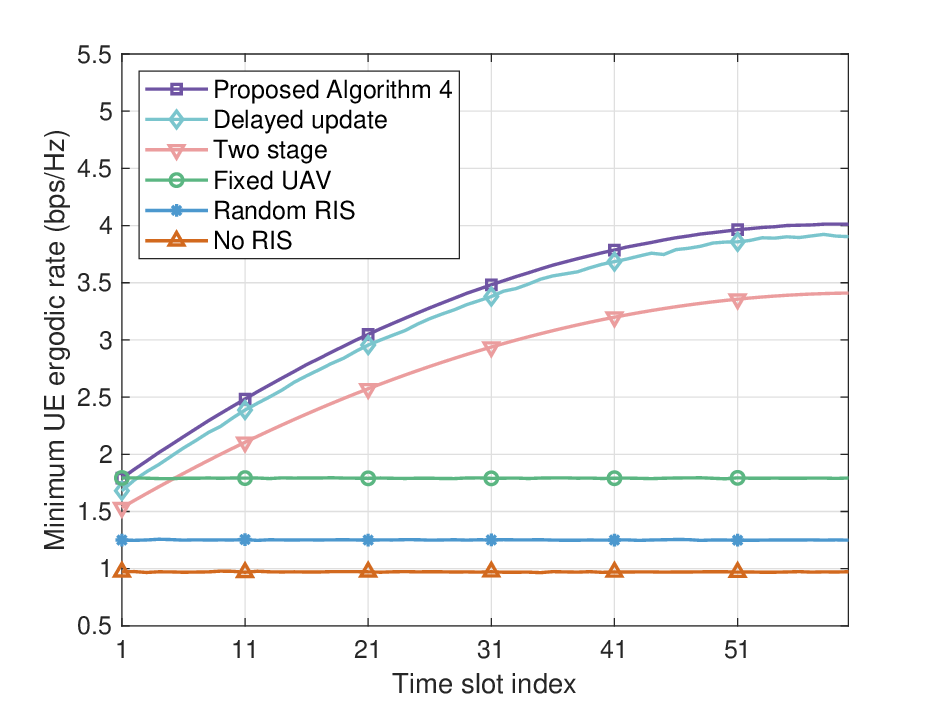}
	\caption{Minimum UE ergodic rate of different schemes.}
	\label{MinR_pre_timeslot}
\end{figure}

Secondly, in Fig. \ref{MinR_pre_timeslot}, we compare the performance of the proposed Algorithm \ref{alg:4} with four baseline schemes from the perspective of minimum UE ergodic rate. In particular, ``Delayed update" represents the case where the system serves the UE using the optimisation results of the previous time slot. ``Fixed UAV" optimizes LEO satellite's beamforming and RIS phase shift while the position of the RIS-UAV is fixed. ``Random RIS" optimizes LEO satellite's beamforming and the UAV trajectory, but phase shift of the RIS is generated randomly. ``No RIS" means that only LEO satellites' beamforming is optimized without the assistance of the RIS-UAV. In addition, ``Two stage" separately optimizes the RIS phase shift by maximizing the channel de-correlation between UEs, and the satellite beamforming by maximizing the sum rate of UEs \cite{hotspot}. It can be seen that the ``Random RIS" scheme does not provide a significant performance gain compared to the ``No RIS" scenario. In contrast, the other three schemes that optimize RIS have a higher rate. This is because the optimization of RIS phase shift can adjust the co-channel interference in multiuser scenarios. Among them, our proposed Algorithm \ref{alg:4} clearly achieves the best performance. Additionally, it can be seen that the minimum ergodic rate remains constant across time slots under the ``No RIS", ``Random RIS", and ``Fixed UAV" schemes, while the minimum ergodic rate optimized by the proposed Algorithm \ref{alg:4} remains increased. This is because the RIS-assisted link fails to provide a higher degree of freedom across time slots due to either absence (No RIS), unoptimized configuration (Random RIS), or static deployment (Fixed UAV). In this case, the optimization of UAV trajectory can coordinate the co-channel interference with more degrees of freedom. Thus, the proposed Algorithm \ref{alg:4} can effectively improve the performance of the LEO satellite constellation. Moreover, it can be seen that the performance loss caused by delayed parameter update is small due to the strong channel correlation between consecutive time slots. This observation confirms the inherent adaptability of the system to minor channel variations over short time intervals.

\begin{figure}[t]
	\centering
	\includegraphics[width=3.4in]{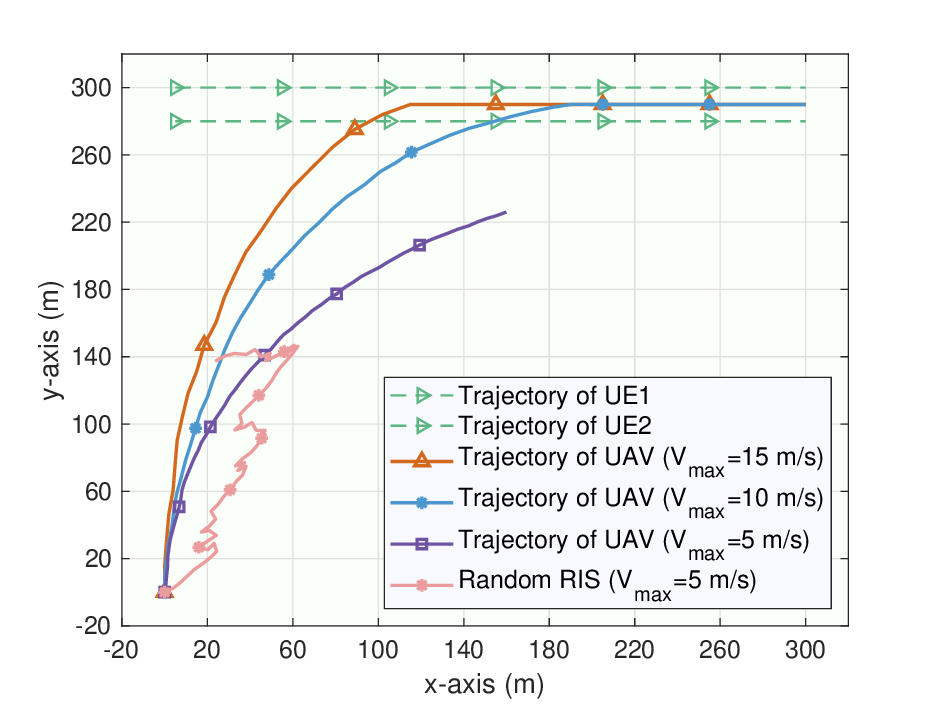}
	\caption{UAV trajectory of different schemes.}
	\label{UAV trajectory}
\end{figure}
Then, we show the optimized UAV trajectories in Fig. \ref{UAV trajectory} under different maximum flight speeds of the RIS-UAV. In particular, we set the number of UEs $K=2$ and these two UEs move along the positive direction of the $x$-axis from the coordinates $\mathbf{q}^{u}_{1} = [0, 300, 0]$ and $\mathbf{q}^{u}_{2}=[0, 280, 0]$ with speed of 5 m/s, respectively. It can be seen that with the RIS design using the proposed Algorithm \ref{alg:4}, the UAV trajectory is directly towards the UEs in order to shorten the distance from the UEs, and when the UAV reaches the optimal position it will fly relatively stationary with the UEs. In contrast, the UAV trajectory does not have a good trend under the ``Random RIS" scheme. This is because signals transmitted from the random RIS may cause more severe interference if the RIS-UAV is closer to the UE. On the contrary, a proper design of the RIS can simultaneously mitigate the interference and enhance the desired signal, resulting in a greater performance gain when the RIS is closer to the UE. This suggests that RIS phase shift design and RIS-UAV trajectory design are integral when the RIS is mounted on the UAV.

\begin{figure}[t]
	\centering
	\includegraphics[width=3.4in]{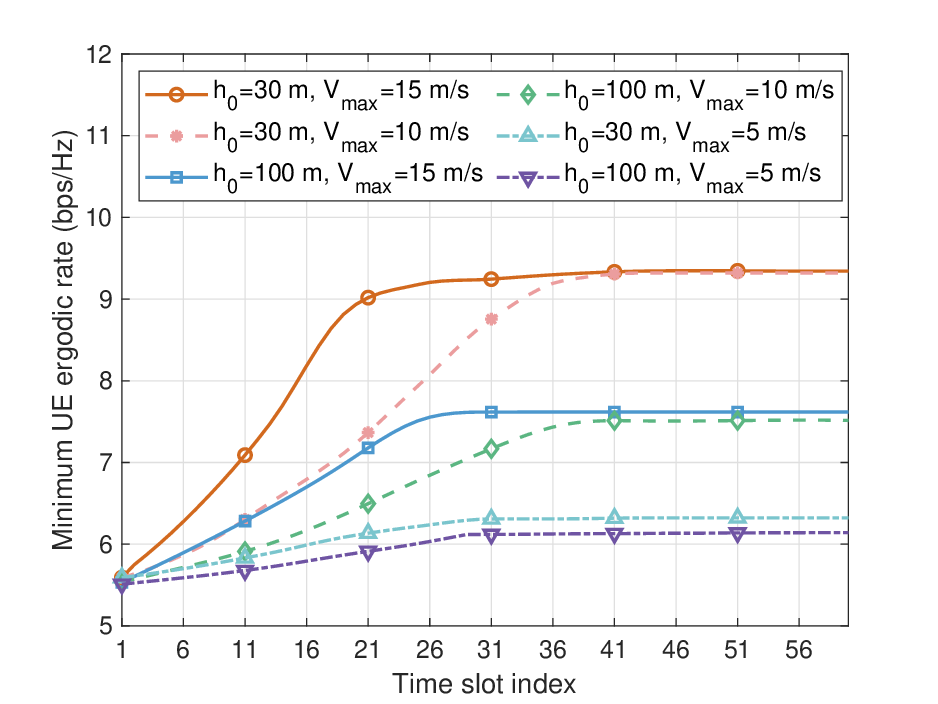}
	\caption{Minimum ergodic rate under different maximum flight speeds and different flight altitudes of UAV.}
	\label{different_speed_different_altitude}
\end{figure}
Fig. \ref{different_speed_different_altitude} examines the effect of different maximum flight speeds and flight altitudes of UAV on system performance. Based on Fig. \ref{UAV trajectory} and Fig. \ref{different_speed_different_altitude}, it is evident that higher maximum speeds of the UAV enable the UAV to reach the optimal position faster, thus enhancing the performance gains achieved by the optimized RIS. When the UAV reaches its optimal position, it maintains synchronised movement with the UEs, ensuring a constant and stable communication quality. In addition, Fig. \ref{different_speed_different_altitude}  shows that increasing the flight altitude of the UAV decreases the system performance. This is due to the fact that higher UAV flight altitudes result in a degradation of the quality of the RIS-assisted link. Therefore, in order to obtain better system performance, the UAV flight altitude should be reduced. However, in practical applications, obstacle avoidance requirements often impose restrictions on the minimum flight altitude. Hence, it is critical to determine an optimal flight altitude that balances performance maximization with safety considerations.

\begin{figure}[t]
	\centering
	\includegraphics[width=3.4in]{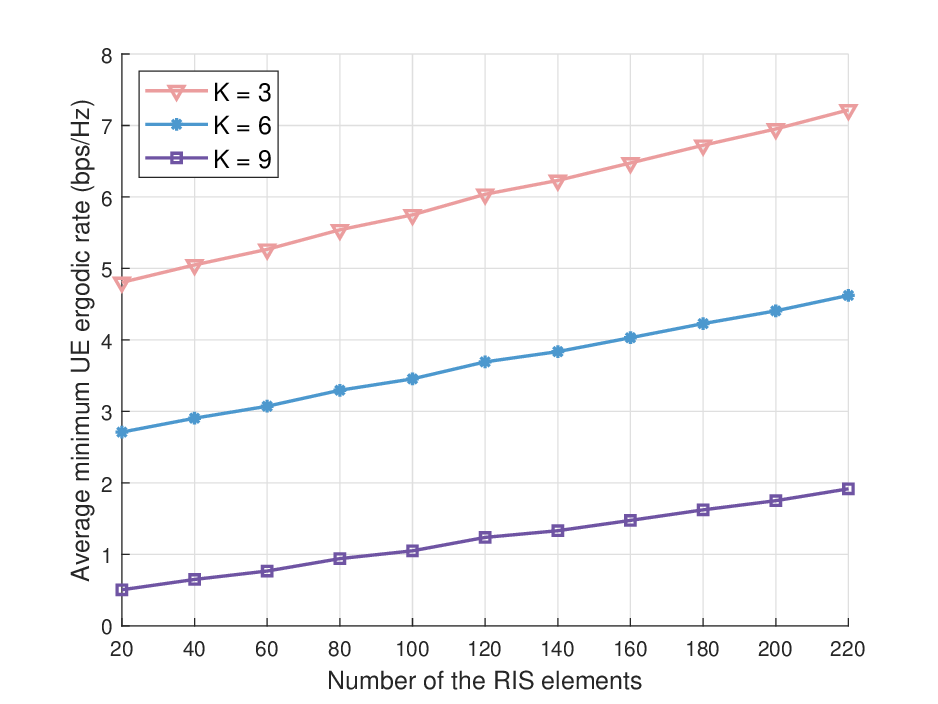}
	\caption{Average minimum UE ergodic rate versus the number of the RIS elements for different numbers of UEs.}
	\label{different_UEs}
\end{figure}
Next, we evaluate the performance of the proposed Algorithm \ref{alg:4} under different numbers of RIS elements for different numbers of UEs $K$ in Fig. \ref{different_UEs}. In particular, the vertical axis represents the average minimum ergodic rate for the duration of the data frame. It is evident that the average minimum ergodic rate decreases rapidly as the number of UE grows within a time-frequency resource block. This is because the transmit power of satellites is divided by more UEs and the co-channel interference between UEs becomes more severe as the number of UEs increases. Therefore, achieving optimal system performance in practical deployments necessitates a holistic resource management approach that harmonizes spatial multiplexing capabilities, UEs scheduling, and resource allocation. In addition, the results reveal that the average minimum UE ergodic rate monotonically increases with the number of RIS elements. This can be attributed to two reasons. Firstly, more elements of RIS result in a larger array gain and hence the signal power is enhanced. Secondly, as the number of elements increases, the RIS has more degrees of freedom to coordinate the multiuser interference. This suggests that RIS can effectively alleviate the impacts of co-channel interference between UEs by adjusting the LoS-dominated satellite-terrestrial channel, which brings a possible solution for high-density UEs under LEO satellite constellation communication. Nevertheless, in order to ensure a stable flight and long endurance of the UAV, the number of elements carried by the RIS-UAV should be limited in practice, even though the size and mass of a single element are very small. Therefore, a proper design of the RIS phase shift with limited transmitting elements is critical for improving the overall performance of LEO satellite constellation communications.

\begin{figure}[t]
	\centering
	\includegraphics[width=3.4in]{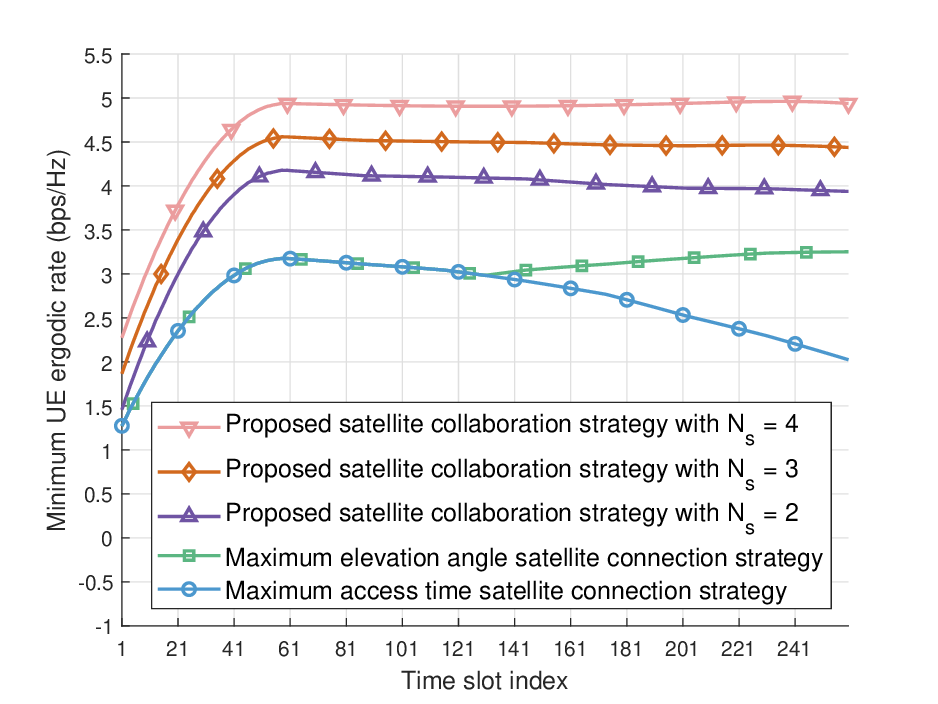}
	\caption{Minimum UE ergodic rate under different satellite connection strategies.}
	\label{DifferentConnection}
\end{figure}

In Fig. \ref{DifferentConnection}, we compare the proposed satellite collaboration strategy under different numbers of collaborative satellites with other satellite connection strategies, i.e. maximum elevation angle satellite connection strategy and maximum access time satellite connection strategy. In particular, the total transmit power budget of LEO satellites at different connection strategies is set the same. As observed in Fig. \ref{DifferentConnection}, the minimum UE ergodic rate initially increases with the time slots before reaching a plateau. This is because the static nature of the UEs leads to a fixed optimal UAV position, which can be asymptotically approached given a sufficient time slot. Compared to the different strategies, it can be seen that the extreme points of the minimum UE ergodic rate increase as the number of collaborating satellites increases. This is because more collaborative satellites provide greater spatial degrees of freedom for optimal satellite beamforming under the same total transmitting power of the collaborating satellites. Whereas, in general, multiple satellites collaboration can result in a larger total transmitting power. In addition, it can be seen that the stability of the minimum UE rate increases as the number of collaborating satellites increases once the UAV reaches the optimal position. In particular, unlike the maximum access time satellite connection strategy, the proposed satellite collaboration strategy shows a minimal reduction in the minimum UE ergodic rate as time slots increase. This stability can be attributed to the balanced distribution of collaborative satellites at varying distances from the UEs and the resulting compensating effect between satellites close to and far from the UEs, which maintains a time-consistent signal power level.

\begin{figure}[t]
	\centering
	\includegraphics[width=3.4in]{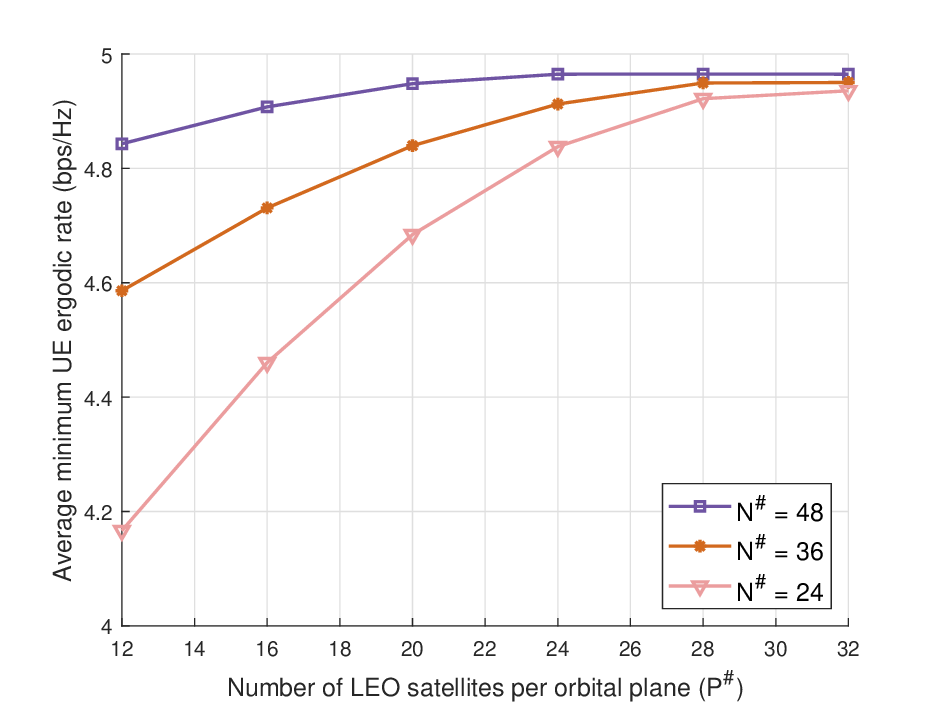}
	\caption{Average minimum UE ergodic rate versus the number of the satellite per orbit for different number of orbit.}
	\label{satellite density}
\end{figure}
Finally, we conduct simulations to evaluate the impact of satellite density by varying both the number of satellites per orbit and the number of orbits, as illustrated in Fig. \ref{satellite density}. It is clear that the average minimum ergodic rate increases as the number of orbits increases and the number of satellites increases. The performance improvement stems primarily from the shorter communication distances enabled by dense LEO satellite deployments, which significantly enhance channel conditions for UE data transmission, as well as the prolonged multi-satellite coverage that contributes to performance gains. This justifies the current industry shift towards larger LEO constellations in next-generation designs, where multi-satellite collaboration will be a fundamental feature of the system. However, it is worth noting that the advantage of dense LEO constellations in terms of system performance diminishes as the density of satellites increases while introducing risks such as orbital collisions, signal interference, and higher costs. Thus, it is crucial to balance coverage requirements, service needs, and cost efficiency when designing satellite constellations.

\section{Conclusion}
In this paper, the deployment of RIS-UAV is advocated to enhance the performance of LEO satellite constellation communications. In order to exploit the potential of RIS-UAV, we formulated the minimum rate of UEs maximization problem and proposed an efficiently iterative algorithm to design the RIS phase shift, UAV trajectory, and LEO satellite beamforming with sCSI. This design effectively accommodated the interaction between the RIS phase shift and the UAV trajectory. Finally, simulation results with various parameters confirmed the effectiveness and superiority of the proposed algorithm. While this work provides a foundational framework, several practical challenges and extensions warrant further investigation. First, the current analysis assumes perfect sCSI estimation, whereas practical implementations face measurement errors and feedback delays due to the passive nature of RIS and long distance propagation. In addition, hardware impairment of RIS is a key factor that degrades the performance of RIS-assisted systems. As a result, it is of interest to achieve robust optimization under imperfect sCSI and hardware impairment conditions in future. Furthermore, the framework considers the deployment of a single RIS-UAV, while the coordination of multiple RIS-UAVs can further enhance the coverage and capacity. Investigating optimal coordination strategies among RIS-UAVs, such as interference management and distributed beamforming, presents an exciting avenue for extending this work.

\appendix
\section{Apendix of derivation of SINR}
\subsection{Derivation of the expression of $\bar{\Gamma}_{k}$}
The $\bar{\Gamma}_{k}$ is the expectation of SINR of the $k$-th UE, which is given by
\begin{align}\label{SINR_Ex1}
    \bar{\Gamma}_{k} = \frac{\mathbb{E}\{\vert\mathbf{f}_{k}^{\text{H}}\mathbf{v}_{k} \vert^2\}}{\mathbb{E}\{\sum\limits_{l\neq k}^{K}\vert\mathbf{f}^{\text{H}}_{k}\mathbf{v}_{l}\vert^2\} + \mathbb{E}\{\sigma^2_{k}\}}.
\end{align}
First, the term $\mathbb{E}\{\vert\mathbf{f}_{k}^{\text{H}}\mathbf{v}_{l} \vert^2\}$ can be unrolled as
\begin{align}\label{Unrolling of SINR}
    &\mathbb{E}\{\vert\mathbf{f}_{k}^{\text{H}}\mathbf{v}_{l} \vert^2\} \notag\\
    = &\mathbb{E}\{\vert ((\bar{\mathbf{h}}_{k}^{\text{H}} + \bar{\mathbf{g}}_{k}^{\text{H}}\boldsymbol{\Theta}\bar{\mathbf{G}}) + \tilde{\mathbf{g}}_{k}\boldsymbol{\Theta}\bar{\mathbf{G}} + \bar{\mathbf{g}}_{k}\boldsymbol{\Theta}\tilde{\mathbf{G}} +\tilde{\mathbf{g}}_{k}\boldsymbol{\Theta}\tilde{\mathbf{G}}\notag \\
    &+ \tilde{\mathbf{h}}_{k})\mathbf{v}_{l} \vert^2\},
\end{align}
where
\begin{align}
\bar{\mathbf{h}}_{k}&=\left[\sqrt{\frac{\kappa_{1,k}L_{1,k}^2}{\kappa_{1,k}+1}}\mathbf{h}^{\text{LoS}}_{1,k}; \cdots ; \sqrt{\frac{\kappa_{S,k}L_{S,k}^2}{\kappa_{S,k}+1}}\mathbf{h}^{\text{LoS}}_{S,k}\right],
\end{align}
\begin{align}
\bar{\mathbf{g}}_{k}&=\sqrt{\frac{\nu_{k}}{\nu_{k}+1}} P_{k}\mathbf{g}_{k}^{\text{LoS}},
\end{align}
\begin{align}
\bar{\mathbf{G}}&=\left[\sqrt{\frac{\kappa^r_1}{\kappa^r_1+1}} L_{1}^r \mathbf{G}^{\text{LoS}}_{1}, \cdots ,\sqrt{\frac{\kappa^r_S}{(\kappa^r_S+1)}} L_K^r \mathbf{G}_{S}^{\text{LoS}}\right],
\end{align}
\begin{align}
\tilde{\mathbf{h}}_{k}&=\left[\sqrt{\frac{L_{1,k}^2}{\kappa_{1,k}+1}}\mathbf{h}^{\text{NLoS}}_{1,k}; \cdots ; \sqrt{\frac{L_{S,k}^2}{\kappa_{S,k}+1}}\mathbf{h}^{\text{NLoS}}_{S,k}\right],
\end{align}
\begin{align}
\tilde{\mathbf{g}}_{k}&=\sqrt{\frac{1}{\nu_{k}+1}} P_{k}\mathbf{g}_{k}^{\text{NLoS}},
\end{align}
\begin{align}
\tilde{\mathbf{G}}&=\left[\sqrt{\frac{1}{\kappa^r_1+1}} L_{1}^r \mathbf{G}^{\text{NLoS}}_{1}, \cdots ,\sqrt{\frac{1}{\kappa^r_S+1}} L_{K}^r \mathbf{G}_{S}^{\text{NLoS}}\right].
\end{align}
Since $\tilde{\mathbf{h}}_{k}$, $\tilde{\mathbf{g}}_{k}$, and $\tilde{\mathbf{G}}$ are mutually independent random variables obeying a zero-mean Gaussian distribution, we have
\begin{align}\label{SINR_Ex2}
    &\mathbb{E}\{\vert\mathbf{f}_{k}^{\text{H}}\mathbf{v}_{l} \vert^2\} \notag\\
    = &\vert (\bar{\mathbf{h}}_{k}^{\text{H}} + \bar{\mathbf{g}}_{k}^{\text{H}}\boldsymbol{\Theta}\bar{\mathbf{G}})\mathbf{v}_{l} \vert^2 + \mathbb{E}\{ \vert \tilde{\mathbf{g}}_{k}\boldsymbol{\Theta}\bar{\mathbf{G}}\mathbf{v}_{l}\vert^2\} +\mathbb{E}\{ \vert\tilde{\mathbf{h}}_{k}\mathbf{v}_{k}\vert^2\} \notag \\
    & +\mathbb{E}\{\vert \bar{\mathbf{g}}_{k}\boldsymbol{\Theta}\tilde{\mathbf{G}}\mathbf{v}_{k}\vert^2  +\mathbb{E}\{\vert\tilde{\mathbf{g}}_{s,k}\boldsymbol{\Theta}\tilde{\mathbf{G}}_s\mathbf{v}_{k}\vert^2\} \}.
\end{align}
Then, by expanding expression (\ref{SINR_Ex2}), we have
\begin{align}
    \mathbb{E}\{ \vert \tilde{\mathbf{g}}_{k}\boldsymbol{\Theta}\bar{\mathbf{G}}\mathbf{v}_{l}\vert^2\} = \frac{1}{\nu_k+1}F_k^2 \Vert \bar{\mathbf{G}}\mathbf{v}_k \Vert^2, \label{70}
\end{align}
\begin{align}
    \mathbb{E}\{\vert \bar{\mathbf{g}}_{k}\boldsymbol{\Theta}\tilde{\mathbf{G}}\mathbf{v}_{k}\vert^2  +\mathbb{E}\{\vert\tilde{\mathbf{g}}_{s,k}\boldsymbol{\Theta}\tilde{\mathbf{G}}_s\mathbf{v}_{k}\vert^2\}\} = \Vert\mathbf{a}_{k}\odot \mathbf{v}_{l} \Vert^2, \label{71}
\end{align}
and
\begin{align}
    \mathbb{E}\{ \vert\tilde{\mathbf{h}}_{k}\mathbf{v}_{k}\vert^2\} = \Vert\mathbf{b}_{k}\odot \mathbf{v}_{l} \Vert^2, \label{72}
\end{align}
where
\begin{align}
\mathbf{a}_{k} = F_k\left[(\sqrt{\frac{M_rL_1^2}{\kappa^r_{1} + 1}}) \mathbf{J}_{N_t\times 1} ; \cdots ; (\sqrt{\frac{M_rL_S^2}{\kappa^r_{S} + 1}} )\mathbf{J}_{N_t\times 1}\right],
\end{align}
and
\begin{align}
\mathbf{b}_{k} = \left[\sqrt{\frac{L_{1,k}^2}{\kappa_{1,k}+1}}\mathbf{J}_{N_t\times 1}; \cdots ; \sqrt{\frac{L_{s,k}^2}{\kappa_{S,k}+1}}\mathbf{J}_{N_t\times 1} \right].
\end{align}
Finally, bringing the unrolled equation (\ref{SINR_Ex2})-(\ref{72}) back to the expression (\ref{SINR_Ex1}), we can obtain the expression of the expectation of SINR (\ref{SINR expression}).

\end{document}